\documentclass[letterpaper]{article} 
\usepackage{aaai2026}  
\usepackage{times}  
\usepackage{helvet}  
\usepackage{courier}  
\usepackage[hyphens]{url}  
\usepackage{graphicx} 
\usepackage{natbib}  
\usepackage{caption} 
\usepackage{algorithm}
\usepackage{algorithmic}

\usepackage{newfloat}
\usepackage{listings}
\DeclareCaptionStyle{ruled}{labelfont=normalfont,labelsep=colon,strut=off} 
\floatstyle{ruled}
\newfloat{listing}{tb}{lst}{}
\floatname{listing}{Listing}
\usepackage{amsfonts}  
\usepackage{amssymb,amsmath}     
\usepackage{bm}
\usepackage{subcaption}

\newcommand\bx{\boldsymbol{x}}
\newcommand\by{\boldsymbol{y}}
\newcommand\bz{\boldsymbol{z}}
\newcommand\ba{\boldsymbol{a}}

\newcommand\bv{\boldsymbol{v}}

\newcommand{\bA}{\boldsymbol{A}}
\newcommand{\bB}{\boldsymbol{B}}
\newcommand{\bI}{\boldsymbol{I}}

\newcommand{\bX}{\boldsymbol{X}}

\newcommand{\setA}{\mathcal{A}}

\newcommand{\setD}{\mathcal{D}}

\newcommand{\setX}{\mathcal{X}}

\newcommand{\setU}{\mathcal{U}}
\newcommand{\setV}{\mathcal{V}}

\newcommand{\bone}{\boldsymbol{1}}  

\newtheorem{Lemma}{Lemma}
\newtheorem{Proposition}{Proposition}
\newtheorem{Definition}{Definition}

\title{A Scalable and Exact Relaxation for Densest $k$-Subgraph via Error Bounds}
\author{
    Ya Liu\textsuperscript{\rm 1}\equalcontrib,
    Junbin Liu \textsuperscript{\rm 1}\equalcontrib, 
    Wing-Kin Ma\textsuperscript{\rm 1},
    Aritra Konar\textsuperscript{\rm 2}\thanks{Corresponding author.}
   }
\affiliations{
    \textsuperscript{\rm 1} The Chinese University of Hong Kong\\
    \textsuperscript{\rm 2} KU Leuven\\
    yaliu@link.cuhk.edu.hk, liujunbin@link.cuhk.edu.hk,  wkma@ee.cuhk.edu.hk, aritra.konar@kuleuven.be
}

\usepackage{bibentry}

\begin{document}

\maketitle

\begin{abstract}
Given an undirected graph and a size parameter $k$, the Densest $k$-Subgraph (D$k$S) problem extracts the subgraph on $k$ vertices with the largest number of induced edges. While D$k$S is NP--hard and difficult to approximate, penalty-based continuous relaxations of the problem have recently enjoyed practical success for real-world instances of D$k$S. In this work, we propose a scalable and exact continuous penalization approach for D$k$S using the error bound principle, which enables the design of suitable penalty functions.
Notably, we develop new theoretical guarantees ensuring that both the global and local optima of the penalized problem match those of the original problem. 
The proposed penalized reformulation enables the use of first-order continuous optimization methods.
In particular, we develop a non-convex proximal gradient algorithm, where the non-convex proximal operator can be computed in closed form, resulting in low per-iteration complexity. 
We also provide convergence analysis of the algorithm.
Experiments on large-scale instances of the D$k$S problem and one of its variants, the Densest ($k_1, k_2$) Bipartite Subgraph (D$k_1k_2$BS) problem, demonstrate that our method achieves a favorable balance between computation cost and solution quality.
\end{abstract}

\begin{links}
    \link{Code}{https://github.com/ly6421/dks-aaai2026}
\end{links}

\section{Introduction}

Dense subgraph discovery (DSD) is a key  graph mining primitive which aims to extract highly interconnected subsets of vertices from a given graph (see \cite{lanciano2024survey,luo2023survey} and the references therein). The problem has garnered considerable attention as it finds myriad applications ranging from mining trending topics in social media \cite{angel2014dense}, discovering complex patterns in gene annotation graphs \cite{saha2010dense}, to detecting fraudsters in e-commerce and financial transaction networks \cite{hooi2016fraudar,li2020flowscope,chen2022antibenford}.

The broad applicability of DSD has resulted in the development of several formulations and algorithms. Notable among these is the Densest Subgraph (DSG) problem \cite{goldberg1984finding}, which extracts the subgraph with the maximum average induced degree.
Although DSG can be solved exactly in polynomial time via maximum flow, it is computationally expensive. A faster greedy peeling algorithm runs in linear time and provides a $0.5$-factor approximation guarantee \cite{charikar2000greedy}. Improved multi-stage extensions of the greedy algorithm were later developed in \cite{boob2020flowless,chekuri2022densest} - these offer improved approximation guarantees for DSG at the same complexity order. A separate line of research  \cite{danisch2017large,harb2022faster,harb2023convergence,nguyen2024multiplicative} has developed algorithms for DSG using tools from convex optimization, building on a linear programming relaxation of DSG due to \cite{charikar2000greedy}. 
A different type of dense subgraph is the max-core of a graph \cite{seidman1983network}, which corresponds to the subgraph that maximizes the minimum induced degree. The greedy algorithm for DSG optimally solves this problem in linear time. 
DSG and the max-core were later unified within a general framework for computing solutions of degree-based density objectives in \cite{veldt2021generalized}.  

A limitation of the DSG and the max-core formulations is that they do not allow the size of the extracted subgraph to be explicitly controlled, which can result in the output being a large subgraph with poor inter-connectivity in real-world graphs, as observed in \cite{tsourakakis2013denser,shin2016corescope}. Indeed, by design, the solution of DSG is maximal in size \cite{harb2023convergence}. Furthermore, as the solution of both problems is unique, there is no flexibility in extracting a denser subgraph of a different size. Arguably, the simplest remedy is to add an explicit size constraint to DSG, which results in the Densest $k$-Subgraph (D$k$S) problem \cite{feige2001dense}. Variation of the size parameter $k$ enables extraction of dense subgraphs of different sizes, thereby allowing users to select a solution with the desired density. 
The flip side of this flexibility is that D$k$S is NP-hard and difficult to approximate in the worst case \cite{manurangsi2017almost}.
However, practical optimization algorithms have been demonstrated to exhibit good performance on real-world instances of the problem \cite{papailiopoulos2014finding,konar2021exploring,lu2025densest,liu2024extremeII}. Our present work advances this line of research by developing a new principled and scalable continuous optimization framework for effectively tackling such ``favorable'' instances of D$k$S. 

Variations of D$k$S which allow extraction of subgraphs of different sizes also exist - these include the Densest at-least-$k$ Subgraph (Dal$k$S) and Densest at-most-$k$ Subgraph (Dam$k$S) problems  \cite{andersen2009finding,khuller2009finding}, and the $f$-densest subgraph problem \cite{kawase2018densest}. 
However, in contrast to D$k$S, these formulations are not guaranteed to span the spectrum of densest subgraphs of {\em every} size.

\noindent 
\textbf{Contributions:} Our main contributions are summarized as follows:

\noindent $\bullet$ We propose a scalable and exact penalization formulation for the D$k$S problem via the error bound principle \cite{luo1996mathematical,CP22}. 
By designing a suitable relaxation of the constraint set of D$k$S and a corresponding error bound function that serves as a penalty, we arrive at a non-convex, non-smooth penalty formulation. 
In the context of penalty methods for constrained optimization, the challenge lies in showing whether a penalty formulation at hand has its globally optimal solution set identical to that of the original problem. In this work, we established not only the equivalence of global optima between our formulations and D$k$S, but also that of local optima.
    
\noindent $\bullet$ To tackle our proposed formulation, we employ a non-convex variant of the classic proximal gradient algorithm \cite{nesterov2018lectures}, comprising a ``forward'' gradient step, and a ``backward'' proximal operator evaluation. A key technical challenge is that evaluating the proximal operator in our formulation entails solving a non-convex problem. Nevertheless, it surprisingly turns out that the global solution of this subproblem can be computed efficiently even in the absence of convexity, resulting in low per-iteration complexity, and thereby endowing the algorithm with scalability. Theoretically, we also show that the proposed algorithm is guaranteed to converge to a critical point of our formulation at a sub-linear rate.

\noindent $\bullet$ We evaluate our method on the D$k$S problem and adapt our method to the Densest ($k_1$, $k_2$) Bipartite Subgraph (D$k_1k_2$BS) problem as well. Extensive experiments on large-scale real-world graphs—with sizes ranging from thousands to millions of nodes—show that our method consistently achieves state-of-the-art performance with low computational cost.

\noindent 
\textbf{Notation: }
Let $\mathbb{R}$ denote the set of real numbers. 
Vectors and matrices are represented by bold lowercase (e.g., $\bx$) and bold uppercase letters (e.g., $\bX$), respectively. 
The $i$th element of $\bx$ is $x_i$, and the $(i,j)$th entry of $\bX$ is $x_{ij}$. 
The transpose of $\bx$ is denoted by $\bx^\top$.
We use $\boldsymbol{0}$ and $\boldsymbol{1}$ to denote the all-zero and all-one vectors, and $\bI$ for the identity matrix. 
The $i$th largest entry of $\bx \in \mathbb{R}^n$ is denoted by $x_{[i]}$, and the \emph{max-$k$-sum} is defined as $S_k(\bx) := x_{[1]} + \cdots + x_{[k]}$.
The notation $\Pi_\setX(\bx)$ denotes the projection of $\bx$ onto the set $\setX$.

\section{Related Work}
Owing to its intrinsic difficulty, D$k$S has been tackled from different perspectives, aiming to strike a good balance between computational cost and solution quality.
The state-of-the-art polynomial-time approximation algorithm for D$k$S due to \cite{bhaskara2010detecting} outputs an $O(n^{1/4 + \epsilon})$-approximation solution in time $O(n^{1/\epsilon})$ for every $\epsilon > 0$, which is quite pessimistic. Alternative schemes include hierarchies of linear programming relaxations \cite{bhaskara2012polynomial} and semidefinite programming (SDP) relaxations ~\cite{feige1997densest,karisch2000solving,bombina2020convex}. While these approaches offer certain optimality guarantees, they are computationally expensive for large-scale problems, mainly due to the fact that these relaxations employ extra variables, thereby increasing complexity. 

More practical approaches for D$k$S include greedy algorithms~\cite{feige2001dense, asahiro2000greedily}; they are computationally efficient but often yield suboptimal solutions due to their inherently myopic nature. Meanwhile, the Truncated Power Method (TPM) of \cite{yuan2013truncated} is a scalable method that converges under certain conditions. 
However, it was found in practice that the subgraphs obtained by TPM can be highly sub-optimal. Another approach is the Spannogram of \cite{papailiopoulos2014finding}, which uses a low-rank factorization of the graph adjacency matrix to approximately solve the D$k$S problem. Using higher rank approximation improves the solution quality, but at the expense of a larger running time, owing to the exponential dependence of the algorithm's complexity on the rank. For example, using just a rank-2 approximation incurs $O(n^3)$ complexity, which limits scalability. 

A separate line of research has blossomed around developing different continuous relaxations of D$k$S followed by application of continuous optimization algorithms - these range from non-convex  \cite{hager2016projection,sotirov2020solving} to convex formulations \cite{konar2021exploring}.  
However, none of these works analyzed whether their respective relaxations are tight or not. 
Closest to our present work are the recent exact penalty-based methods of ~\cite{liu2024cardinality, lu2025densest, liu2024extreme, liu2024extremeII}, which have demonstrated promising performance in addressing large-scale D$k$S problems. The general principle behind these approaches is to relax the combinatorial constraint set of D$k$S to arrive at a continuous optimization problem and employ penalty terms to enforce the desired combinatorial constraints. Through different lines of reasoning, these works establish equivalence between the globally optimal solutions of their formulations and those of D$k$S.
First-order continuous optimization algorithms are adopted for the resulting formulations, leading to low computational cost. 
Although our approach falls in the above category of methods, it offers several benefits over the prevailing state-of-the-art (see Contributions and {\bf (D1)-(D2)}) in the next section.

\section{Background and Preliminaries}
\label{sec_bg}

Let \( \mathcal{G} = (\mathcal{V}_n, \mathcal{E}_m) \) be an undirected, unweighted graph with \( \mathcal{V}_n \) being the vertex set containing $n$ vertices and $\mathcal{E}_m$ being the edge set containing $m$ edges. 
Given a size parameter $k < n$, the D$k$S problem aims to identify a vertex set $\mathcal{S}_k\subseteq\mathcal{V}_n$ with $k$ vertices such that the number of edges between vertices in $\mathcal{S}_k$ is maximized.
Formally, the problem can be expressed as 

\begin{equation}
\label{eq_dks}
\begin{aligned}
\max_{\bx \in \mathbb{R}^n} \quad  \bx^\top \bA \bx,\quad \text{s.t.}\   \bx \in \mathcal{U}_k^n
\end{aligned}
\end{equation}
where \( \bA \in \mathbb{R}^{n \times n} \) denotes the adjacency matrix of $\mathcal{G}$, and $\setU^n_k:= \{ \bx\in\{0,1\}^n\mid \bx^\top \bone =k\}$ is a set of binary vectors with $k$ nonzero entries. We term the set $\setU^n_k$ as the \emph{selection vector set}. 
In this work, we adopt a continuous optimization approach for the purpose of obtaining good approximate solutions to \eqref{eq_dks}. The theoretical underpinnings are based on the concept of \emph{exact penalization}, which is formally defined below.

\begin{Definition}
\label{def_exact_pen}
{\bf(Exact Penalization)} Consider the following  constrained optimization problem:
\begin{equation}\label{eq_ep_ori}
    \min_{\bx\in\mathcal{A}} f(\bx), 
\end{equation}
where $f: \mathcal{D}\rightarrow \mathbb{R}$ is the objective function defined on a domain $\mathcal{D}\subseteq \mathbb{R}^n$, and $\mathcal{A}\subseteq \mathcal{D}$ is a non-empty constraint set.
Assume that the optimal solution set of \eqref{eq_ep_ori} is non-empty.
Now consider the penalization formulation:
\begin{equation}\label{eq_ep_pen}
    \min_{\bx\in\mathcal{X}} f(\bx)+\lambda h(\bx), 
\end{equation}
where 
\(\mathcal{X} \subseteq \mathcal{D} \) is a constraint set such that $\mathcal{A}\subseteq\setX$, $h:\mathbb{R}^n\rightarrow \mathbb{R}$ is a penalty function for enforcing the constraint satisfaction of $\setA$, and \( \lambda > 0 \) is a given penalty parameter.
If there exists a finite \( \lambda \) such that the optimal solution set of \eqref{eq_ep_pen} is the same as that of \eqref{eq_ep_ori}, then~\eqref{eq_ep_pen} is called an exact penalization of~\eqref{eq_ep_ori}.
\end{Definition}

An instance of exact penalization for the D$k$S problem is the following formulation 
\begin{equation}
\label{eq_dks_expp}
\begin{array}{ll}
\displaystyle \min_{\bx \in \mathbb{R}^n}   -\bx^\top \bA \bx - \lambda\|\bx\|^2_2,\ 
\mathrm{s.t.} \  \bx \in \text{conv}\left(\setU_k^n\right),
\end{array}
\end{equation}
where $\text{conv}\left(\setU_k^n\right)=\{ \bx\in[0,1]^n\mid \bx^\top \bone =k\}$ is the convex hull of $\setU_k^n$ and $\lambda>0$ is the penalization parameter.
The \emph{exact penalization} property of \eqref{eq_dks_expp} was shown from different perspectives.
The study in~\cite{liu2024extreme,liu2024extremeII} showed exactness through both the concave minimization principle and the error bound principle~\cite{luo1996mathematical, CP22}. 
The study in~\cite {lu2025densest} leveraged an extension of the Motzkin-Straus theorem~\cite{motzkin1965maxima} to prove exactness. 
Formulation \eqref{eq_dks_expp} is a non-convex optimization problem with a convex constraint, and hence, it can be readily tackled by first-order optimization algorithms. 
The work~\cite{lu2025densest} employed the Frank-Wolfe algorithm, and the paper~\cite{liu2024extremeII} adopted projected gradient descent combined with majorization-minimization.
However, prior studies working with problem \eqref{eq_dks_expp} have their drawbacks.

{\bf (D1):} As the optimization landscape of \eqref{eq_dks_expp} is non-convex, optimization algorithms may converge to local minima, which are not guaranteed to be feasible for D$k$S; i.e., they may fall outside the constraint set $\mathcal{U}_{k}^n$. Therefore, establishing equivalence between the sets of local optima of the penalized and original formulations is of practical value.
While the work \cite{lu2025densest} proves global optimality equivalence, the relationship between local optima was not addressed. 
The result in \cite{liu2024extreme,liu2024extremeII} 
essentially suggests that formulation \eqref{eq_dks_expp}
has both global and local solution equivalence.
The limitation of this prior result is however that we need the convex hull $\text{conv}(\setU_n^k)$ as the constraint set, which means that the result may no longer apply if we consider a different relaxation set.

{\bf (D2):} As mentioned, the existing studies consider the convex hull of $\mathcal{U}_k^n$ as the relaxation set.
We are concerned with the computational aspects arising from the use of $\text{conv}(\setU_n^k)$ as the constraint set.
The projected gradient methods require projecting onto $\text{conv}(\mathcal{U}_k^n)$, which does not have a closed-form solution and requires a bisection search with complexity $\mathcal{O}(n \log n)$~\cite{konar2021exploring}. 
As the problem size increases, the associated computational cost can become non-negligible.
The Frank-Wolfe Algorithm \cite{lu2025densest} has a per iteration cost of $\mathcal{O}(n \log k)$.
In practice, however, it was noted that the Frank-Wolfe Algorithm can be slow compared to the proximal gradient methods (which will be reflected in our numerical results).
This motivates the exploration of alternative, easier-to-work-with relaxation sets with appropriate penalization functions that maintain exactness for both global and local optimal solution sets and support efficient algorithm design.\\
\noindent{\bf The Error Bound Principle:} The error bound principle \cite{luo1996mathematical,CP22} is a general optimization technique that can be leveraged for exact penalization. 
The same principle was also used to develop the exact penalization \eqref{eq_dks_expp} for D$k$S in ~\cite{liu2024extreme,liu2024extremeII}.
However, as we will see later, our invocation of error bounds results in a penalization formulation that differs significantly from that of ~\cite{liu2024extreme,liu2024extremeII}.

To introduce the error bound principle, we need the notion of error bound functions.
\begin{Definition}
\label{def_errorbound} {\bf(Error Bound Function)}
Given a set $\setX \subseteq \mathbb{R}^n$ and a set $\setA \subseteq \setX$, a function $\psi: \mathbb{R}^n \rightarrow \mathbb{R}$ is said to be an error bound function of $\setA$ relative to $\setX$ if  
\begin{equation} \label{eq:err_bnd_def_a}
\begin{array}{ll}
\mathrm{dist}(\bx, \setA) \leq \psi(\bx), & \quad \forall \bx \in \setX,
\end{array}
\end{equation}
\begin{equation} \label{eq:err_bnd_def_b}
\begin{array}{ll}
\psi(\bx) = 0, & \quad \forall \bx \in \setA,
\end{array}
\end{equation}
where {\rm{dist}} is defined as
\[
\mathrm{dist}(\bx, \mathcal{A}) := \inf_{\by \in \mathcal{A}} \| \bx - \by \|_2.
\]
\end{Definition}
An error bound function provides an upper bound on the distance from a point to the target set $\mathcal{A}$. 
Intuitively, when used as a penalty function, it promotes feasibility by encouraging solutions to lie within the target set $\mathcal{A}$. 
The following lemma formalizes this intuition, showing that error bound functions can yield exact penalization.
\begin{Lemma}
	{\bf 
	(Theorem 2.1.2 in \cite{luo1996mathematical}, or 
    {Proposition 9.1.1} in \cite{CP22})	
	}
	\label{lem:ep2}
	Let $\setD \subseteq \mathbb{R}^n$.
	Let $\setA \subseteq \setD$ be a non-empty closed set.
	Let $f: \setD \rightarrow \mathbb{R}$ be a function that is $K$-Lipschitz continuous on some set $\setX \subseteq \setD$, with $\setA \subseteq \setX$.
	Suppose that 
	\begin{equation}
\label{eq:gen_prob2}
	\min_{\bx \in \setA} f(\bx)
	\end{equation} 
	has an optimal solution.
    Let $\psi$ be an error bound function of $\setA$ relative to $\setX$.
	Given any scalar $\lambda > K$, the following problem
	\begin{equation}
\label{eq:gen_prob2_ep}
	\min_{\bx \in \setX} f(\bx) + \lambda \, \psi(\bx)
	\end{equation}  
	is an exact penalization of problem \eqref{eq:gen_prob2}.
\end{Lemma}
This lemma establishes that for problems with Lipschitz continuous objective functions, exact penalization can be achieved by constructing a valid error bound function \( \psi \) and selecting a sufficiently large but finite penalty parameter  \( \lambda \). 
Notably, many functions—including the objective in the D$k$S formulation \eqref{eq_dks}—are Lipschitz continuous over compact sets.
An important aspect of this approach is that the error bound function is intrinsically tied to the choice of the relaxation set $\setX$. 
Unlike traditional penalty methods such as the proximal distance method~\cite{keys2019proximal} and the quadratic penalty method~\cite{nocedal2006numerical}, which reformulate the original problem as an unconstrained minimization with penalty terms, the error bound approach emphasizes the design of both the relaxation set and the penalty function. 
This often results in more structured and tractable penalization formulations.
As mentioned previously, the method in~\cite{liu2024extreme,liu2024extremeII} is one such example.
It chooses $\text{conv}(\setU_k^n)$ as the relaxation set. 
The objective function $f(\bx)=-\bx^\top\boldsymbol{A}\bx$ is $2\sqrt{k}\|\boldsymbol{A}\|_2$-Lipschitz continuous over the set $\text{conv}(\setU_k^n)$. 
It can then be shown that $\psi(\bx)=2(k-\|\bx\|_{2}^2)$ is an error bound function of $\setU_k^n$ relative to its convex hull $\text{conv}(\setU_k^n)$.
This leads to the formulation \eqref{eq_dks_expp}.

\section{Proposed Approach}
\label{sec_ourmethod}
We develop a new penalization approach for tackling D$k$S by leveraging the error bound principle. For this task, we need two ingredients: (i) a relaxation set of the constraints $\setU_k^n$, and (ii) a suitable error bound function.
For the first part, instead of relaxing $\setU_k^n$ to its convex hull $\text{conv}(\setU_k^n)$, we relax it to the hypercube $[0, 1]^n$, which admits a simpler description.
The next step is to construct an algorithmically tractable error bound function, which serves as a penalty to encourage points in $[0, 1]^n$ to move toward the selection vector set $\setU^n_k$. 
To this end, we have the following result.
\begin{Lemma} \label{thm_eb2_Unk_01}
For any \( \bx \in [0, 1]^n  \), the following inequality holds:
\begin{equation}
\mathrm{dist}(\bx, \setU^n_k) \leq \psi(\bx):= k + \bone^\top \bx - 2 S_k(\bx),
\label{eq_prop_dist01}
\end{equation}
where \( S_k(\bx) = \sum_{i=1}^k x_{[i]} \), and \( x_{[1]} \geq x_{[2]} \geq \dots \geq x_{[n]} \) denotes the sorted components of \( \bx \) in descending order. 
Also, \( \psi(\bx) = 0 \) for \( \bx \in \setU^n_k \). 
The function \( \psi(\bx) \) is an error bound function of \( \setU^n_k \) relative to \( [0, 1]^n \).
\end{Lemma}

{\em Proof of Lemma \ref{thm_eb2_Unk_01}:} \
Let $ \bx\in[0, 1]^n$. 
Without loss of generality, assume $ x_1\geq x_2\geq \dots \geq x_n$. 
Let $\by = \Pi_{\mathcal{U}^n_k}(\bx)$ denote the projection of $\bx$ onto $\mathcal{U}^n_k$. 
Then,  $ \by = (1,\dots, 1, 0,\dots,0) $ with the first $k$ elements being $1$.
The distance function can be upper bounded as:
\begin{equation}
\begin{aligned}
\mathrm{dist}(\bx, \mathcal{U}^n_k) & = \|\bx - \by\|_2 \leq \|\bx - \by\|_1 \\
& = \displaystyle \sum_{i=1}^k (1 - x_i) + \sum_{i=k+1}^n x_i \\
& = k + \displaystyle \sum_{i=1}^n x_i - 2\sum_{i=1}^k x_i\\
& = k + \bone^\top \bx - 2 S_k(\bx).  
\end{aligned}
\label{eq_dist_infty}
\end{equation}
When $\bx\in\setU_k^n$, we have 
\[
\mathrm{dist}(\bx, \mathcal{U}^n_k) = k + \bone^\top \bx - 2 S_k(\bx)=k + k - 2 k=0.\]
The proof is complete.
\hfill $\blacksquare$
\medskip

Based on the error bound function $\psi(\bx)$ in (\ref{eq_prop_dist01}), we propose the following penalized reformulation of the D$k$S problem \eqref{eq_dks}:
\begin{equation}
\min_{\bx\in [0, 1]^n} F_\lambda(\bx):= \underbrace{-\bx^\top\bA\bx}_{=f(\bx)} + \lambda \underbrace{\Big[ \bone^\top \bx -2S_k(\bx) \Big]}_{:= {h}(\bx)}.
\label{eq_dks_01}
\end{equation}
It is straightforward to verify that the objective function $f(\cdot)$ is $2\sqrt{n}\|\bA\|_2$-Lipschitz continuous over $[0, 1]^n$.  Consequently, exact penalization holds by Lemma \ref{lem:ep2}.
More importantly, formulation \eqref{eq_dks_01} also has the local optimal solution set equivalence, as stated in the following proposition.
\begin{Proposition}
\label{prop:ep}
Consider the D$k$S problem (\ref{eq_dks}) and its penalized reformulation \eqref{eq_dks_01}.
Given any scalar $ \lambda > 2\sqrt{n}\|\bA\|_2 $, problem \eqref{eq_dks_01} is an equivalent formulation of problem \eqref{eq_dks} in the sense that both the {\bf global} and {\bf local} optimal solution sets of \eqref{eq_dks_01} are the same as the corresponding sets of \eqref{eq_dks}. 
\end{Proposition}
{The proof of Proposition \ref{prop:ep} is relegated to Part A of the supplementary material.}
Compared to Lemma~\ref{lem:ep2}, Proposition~\ref{prop:ep} extends the result from the equivalence of global optimal solution sets to the equivalence of local optimal solution sets \textit{without any} additional assumptions. This offers a practical guarantee: if an algorithm designed to solve~\eqref{eq_dks_01} converges to a local minimizer, then this solution lies in \( \mathcal{U}_k^n \) and is also a local minimizer of the original problem~\eqref{eq_dks}.
Moreover, Proposition~\ref{prop:ep} is not limited to the D$k$S problem; it is applicable to a broader class of optimization problems constrained by the selection vector set, provided that the objective function is Lipschitz continuous on $[0, 1]^n$.\\
\noindent{\bf A Non-Convex Proximal Gradient Algorithm:}
In this subsection, we develop a non-convex, first-order algorithm for solving \eqref{eq_dks_01}.
Since the cost function of \eqref{eq_dks_01} has a composite form comprising both differentiable and non-differentiable terms, we consider the proximal gradient method (PGM), whose iterations are given by \cite{nesterov2018lectures}:
\begin{equation}
\label{eq:pgd}
\begin{aligned}
\bz^\ell &= \bx^\ell + \gamma_\ell(\bx^\ell-\bx^{\ell-1})\\
\bx^{\ell+1} &= \text{prox}_{[0, 1]^n, \eta_\ell \lambda h}( \bz^\ell - \eta_\ell \nabla f(\bz^\ell) ),   
\end{aligned}
\end{equation}  
where $l=0,1,...,$ index iterations, $\bx^0=\bx^{-1}$, $\{\eta_\ell\}_{\ell\geq 0}$ is a step size sequence, $\{\gamma_\ell\}_{\ell\geq 0}$ is an extrapolation sequence and we adopt the FISTA extrapolation \cite{beck2009fast};
$\mathrm{prox}_{[0,1]^n,\eta_\ell  \lambda h}$ denotes the proximal operator and is defined as
\begin{equation}\label{eq_prox_op}
  \mathrm{prox}_{[0,1]^n,\eta_\ell  \lambda h}(\bz)
= \displaystyle \operatorname*{arg\,min}_{\bx \in [0,1]^n} \frac{1}{2} \|\bz - \bx\|_2^2 +\eta_\ell  \lambda h(\bx).  
\end{equation}
Note that we add extrapolation in the PGM, which in practice was found to lead to faster convergence.
Computing the gradient of $f(\cdot)$ in each iteration entails performing a sparse-matrix vector multiplication, which can be accomplished in time $O(m)$ where $m$ is the number of edges. 
A critical consideration for the PGM is whether the proximal operator can be computed efficiently.
In our case, with $h(\bx)=\bone^\top \bx - 2 S_k(\bx)$, problem  \eqref{eq_prox_op} is not guaranteed to be convex for every $\lambda$.
Nevertheless, we show that despite the absence of convexity, a global optimum can still be computed in an efficient manner.
\begin{Proposition}
\label{lem_dks_prox}
Consider the following non-convex optimization problem
\begin{equation}
\begin{array}{rl}
\min\limits_{\bx \in [0,1]^n} \frac{1}{2} \|\bz - \bx\|_2^2 + \mu (\bone^\top \bx - 2 S_k(\bx)),
\end{array}
\label{eq_P}
\end{equation}
where $\mu>0$ is given.
Without loss of generality, assume \( z_1 \geq z_2 \geq \dots \geq z_n \). 
The following vector
\begin{equation*}
\bx^* = \left( [z_1 + \mu]_0^1, \dots, [z_k + \mu]_0^1, [z_{k+1} - \mu]_0^1, \dots, [z_n - \mu]_0^1 \right)
\label{eq_prox_sln}
\end{equation*}
is a global optimal solution to problem (\ref{eq_P})
where \( [a]_0^1 := \min(1, \max(0, a)) \) clips the variable to lie within $[0, 1]$.
\end{Proposition}
\medskip
The proof is provided in Part B of the supplementary material.
To evaluate the proximal operator, it suffices to identify the top-\( k \) entries of \( \bz \), which can be done in \( \mathcal{O}(n \log k) \) time using a min-heap.
This is generally faster than the \( \mathcal{O}(n \log n) \) cost required to project onto \( \mathrm{conv}(\mathcal{U}_k^n) \)~\cite{liu2024extremeII,konar2021exploring}, thereby enhancing scalability, particularly when identifying small subgraphs within extremely large graphs.
This advantage will be demonstrated in the section on numerical results.

In addition to its low per-iteration computational cost, the proposed algorithm is guaranteed to converge to a critical point of Problem~\eqref{eq_dks_01} at a sublinear rate.
Here, a point $\bx$ is defined as a critical point if
\begin{equation}
\boldsymbol{0} \in \partial F_{\lambda}(\bx)= \nabla f(\bx) + \lambda\, \partial h(\bx),
\end{equation}
where $\partial h(\bx)$ denotes the limiting subdifferential of $h$ at $\bx$~\cite{rockafellar2009variational,li2020understanding}. 
The following convergence result is established.
{
\begin{Proposition}
\label{prop_dks_prox_convergence}
	Consider the PGM \eqref{eq:pgd} for Problem \eqref{eq_dks_01}. 
     Suppose the step sizes satisfy $c_1L_f\leq 1/\eta_\ell\leq c_2L_f$ with $1<c_1\leq c_2<\infty$, and the extrapolation weights satisfy $0\leq\gamma_\ell< \bar \gamma<1$ with $\bar\gamma=(c_1-1)/(2+2c_2)$.
Then, the sequence \(\{\bx^{\ell}\}_{\ell \geq 0}\) generated by the PGM exhibits a sublinear convergence rate property
	\begin{equation}
	\min_{\ell = 0, \dots, J} {\rm dist}(\boldsymbol{0}, \partial F_\lambda(\bx^{\ell+1})) \leq \sqrt{C/(J+1)},
	\end{equation}
	where
    \[
    C=\frac{64(1+c_2^2)(c_1-1)(F_\lambda(\bx^0) - F_\lambda^*)\|\boldsymbol{A}\|_2}{(c_1-1)^2-4\bar\gamma^2(c_2+1)^2}
    \]
    and  $F_\lambda^*=\min\limits_{\bx\in[0,1]^n}F_\lambda(\bx)$.
\end{Proposition}
We provide the proof in Part C of the supplementary material.
It should be noted that convergence analysis of the PGM with extrapolation and for a non-convex objective function was previously considered, e.g., in \cite{xu2017globally}.
There, the convergence rate result requires the Kurdyka--{\L}ojasiewicz (KL) property with the objective function $f$. 
The present convergence result differs in that we do not use the KL property, and we only use the Lipschitz continuous gradient property.
We point out that the proposed non-convex PGM and its convergence analysis are applicable to other problems with the selection vector constraint, provided that the cost function has Lipschitz continuous gradients, and hence, it may prove to be of broader interest.}

In practice, it has been observed that the algorithms may get stuck at poor local minima if the penalty parameter $\lambda$ is set too large from the outset~\cite{shao2020binary, zaslavskiy2008path}.
To address this, $\lambda$ is typically annealed from a small initial value to a larger one, eventually reaching the regime of exact penalization. 
We adopt this strategy in our implementation as well.
The entire algorithm is summarized in Algorithm \ref{alg_whole}.

\begin{algorithm}[!t]
\caption{The Proposed Algorithm for Solving (\ref{eq_dks_01})}
\label{alg_whole}
\textbf{Input}: Initial point \( \bx^0 \), initial penalty \( \lambda_0 \), an extrapolation sequence $\{\gamma_\ell\}_{\ell\geq 0}$, a step size sequence $\{\eta_\ell\}_{\ell\geq 0} $\\
\textbf{Output}: Final solution \( \bx^* \)
\begin{algorithmic}[1]
\STATE \(\bx^{-1}  =\bx^0  \)
\FOR{\( \ell = 0, 1, \dots \)}
       \STATE \({\bz}^{\ell} \leftarrow \bx^{\ell} + \gamma_\ell(\bx^{\ell}-\bx^{\ell-1}) \)
       \STATE \( {\bx}^{\ell+1} \leftarrow \mathrm{prox}_{[0,1]^n,\, \eta_\ell\lambda_\ell h}({\bz}^\ell - \eta_\ell \nabla f({\bz}^\ell) ) \)
        \IF {a penalty update condition is met}
            \STATE increase the penalty parameter: \( \lambda_{\ell+1} > \lambda_\ell \)
        \ENDIF
    \IF {a stopping criterion is met}
        \STATE \textbf{break}
    \ENDIF
\ENDFOR
\STATE \textbf{return} \( \bx^* = \bx^{\ell+1} \)
\end{algorithmic}
\end{algorithm}

\noindent{\bf Adaptation to the D$k_1k_2$BS Problem:}
The proposed approach can be easily adapted to the D$k_1k_2$BS problem, a variant of the D\(k\)S problem.
Given an undirected bipartite graph $\mathcal{G} = (\mathcal{V}_1, \mathcal{V}_2, \mathcal{E}_m)$, where \(\mathcal{V}_1\) and \(\mathcal{V}_2\) are disjoint vertex sets with sizes \(n_1 = |\mathcal{V}_1|\) and \(n_2 = |\mathcal{V}_2|\), respectively, such that \(n = n_1 + n_2\), 
the edge set \(\mathcal{E}_m\) contains connections exclusively between nodes in \(\mathcal{V}_1\) and \(\mathcal{V}_2\).
The goal of the D$k_1k_2$BS problem is to select $k_1<n_1$ vertices from $\mathcal{V}_{1}$ and $k_2<n_2$ vertices from $\mathcal{V}_{2}$ such that the number of edges between the selected vertices---i.e., edges connecting one vertex in $\mathcal{V}_{1}$ to one in $\mathcal{V}_{2}$---is maximized.
This problem also finds wide applications.
For example, in recommendation systems, interactions between users and items can be naturally modeled as a bipartite graph, where identifying dense subgraphs helps reveal strong user-item affinities.

Formally, the D$k_1k_2$BS problem can be formulated as:
\begin{equation}
\label{eq_dkbs}
\begin{array}{ll}
\displaystyle \min_{\bx \in \mathbb{R}^{n_1},\ \by \in \mathbb{R}^{n_2}} -\bx^\top \bB \by,\quad
\mathrm{s.t.} \  \bx \in \setU_{k_1}^{n_1},\  \by \in \setU_{k_2}^{n_2},
\end{array}
\end{equation}
where $\bB\in \mathbb{R}^{n_1\times n_2}$ denotes the biadjacency matrix for a bipartite graph, and \( \bx \in \{0,1\}^{n_1} \), \( \by \in \{0,1\}^{n_2} \) are selection vectors.
Based on our error bound result, we can reformulate the D$k_1k_2$BS problem as
\begin{equation}
\label{eq_dkbs_ep}
\begin{array}{ll}
\displaystyle \min_{\bx,\ \by} & -\bx^\top \bB \by+\lambda\big(\bone^\top\bx+\bone^\top\by-2S_{k_1}(\bx)-2S_{k_2}(\by)\big), \\[1ex]
\mathrm{s.t.} & \bx \in [0, 1]^{n_1},\quad \by \in [0, 1]^{n_2}.
\end{array}
\end{equation}
The equivalence between \eqref{eq_dkbs_ep} and \eqref{eq_dkbs}, in the sense of Proposition \ref{prop:ep}, holds.
This is because the error bounds for two disjoint sets can be combined to yield an error bound for their Cartesian product ~\cite{liu2024extreme}.
For implementation, we use the proposed proximal gradient method where the proximal operator is essentially the same as before.
More details are provided in  Part E of the supplementary material.

\section{Numerical Results}
\label{sec_results}
In this section, we present empirical evaluations of our proposed algorithm on the D$k$S problem and the D$k_1k_2$BS problem. 
We term our method, the exact penalization formulation solved by the proximal gradient method, as EP-Prox.
\subsection{The D$k$S Problem}
\noindent{\bf Implementation:} 
We implement our method using Algorithm~\ref{alg_whole}.
{The initial penalty parameter is set to \( \lambda_0 = 10^{-10} \). 
The algorithm terminates when either \( \| \bx^{\ell+1} - \bx^\ell \|_2^2 \leq 10^{-11} \), or the number of iterations exceeds $100$. The penalty parameter \( \lambda \) is updated by \( \lambda_{\ell+1} = 20 \lambda_\ell \) when either \( \| \bx^{\ell+1} - \bx^\ell\|_2 / \| \bx^{\ell+1} \|_2 < 0.5 \) or $10$ iterations have passed since the last update.

\noindent{\bf Baselines:} We used six widely-used baselines for D$k$S: (i) the greedy method \cite{feige2001dense}; (ii) the truncated power method (TPM) \cite{yuan2013truncated}; (iii) rank-$ 1 $ binary principal component (Rank-1 PC) approximation \cite{papailiopoulos2014finding} \footnote{\url{https://github.com/mitliagkas/spannogram}};
(iv) the Lov\'{a}sz relaxation via Linearized-ADMM (Lov\'{a}sz)  \cite{konar2021exploring} \footnote{\url{https://github.com/luqh357/DkS-Diagonal-Loading}}; (v) the extreme point pursuit (EXPP) method \cite{liu2024extreme, liu2024extremeII}; (vi) the Frank-Wolfe (FW) method \cite{lu2025densest} \footnote{\url{https://github.com/luqh357/DkS-Diagonal-Loading}}.
We used our own implementation of (i) and (ii).
All the methods are initialized with the scaled all-one vector, $\bx^0 = \bone/n$.

\noindent{\bf Datasets:} We evaluate the methods on a diverse set of real-world graphs obtained from \cite{snapnets}.
The detailed statistics of the datasets are summarized in Table \ref{table_stat_dks}.
\begin{table}[!t]
	\centering
	\begin{tabular}{c|c|c}
		\hline
		Datasets & $n$ & $m$ \\
		\hline\hline
		HepTh & 9,877 & 25,998 \\
		\hline
        CondMat & 23,133 &	93,497 \\
		\hline 
		DBLP & 317K & 1M \\
		\hline
		roadNet & 1.9M & 2.7M \\
		\hline
		Talk & 2.4M & 5M \\
        \hline
        LiveJournal1 & 4.8M & 69M\\
		\hline 
	\end{tabular}
	\caption{Statistics of the datasets, where $ n $ represents the number of nodes and $m$ represents the number of edges.}
	\label{table_stat_dks}
\end{table}
Each dataset was preprocessed by symmetrizing directed arcs (if the original dataset is a directed graph), removing all self loops, and extracting the largest connected component.
To assess solution quality, we use the standard edge density metric: 
\begin{equation}
{\rm edge~density }= \bx^\top\bA\bx/(k(k-1)),
\label{eq_edge_density}
\end{equation}
where $\bx\in \setU_{k}^n$ is the binary indicator vector representing the selected subgraph.
Note that $k$ is a fixed constant, and the numerator corresponds to the original objective function that we want to maximize in problem~\eqref{eq_dks}. 
Consequently, a larger edge density value reflects a higher-quality solution.

\begin{figure}[!t]
\centering
\begin{subfigure}[b]{0.48\columnwidth}
  \centering
  \includegraphics[width=\linewidth]{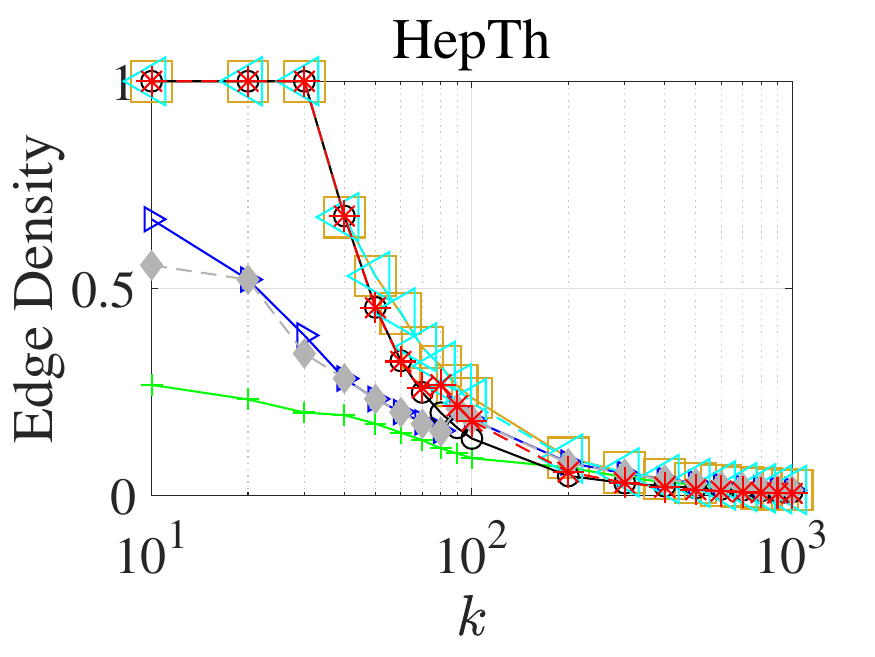}
\end{subfigure}
\hfill
\begin{subfigure}[b]{0.48\columnwidth}
  \centering
  \includegraphics[width=\linewidth]{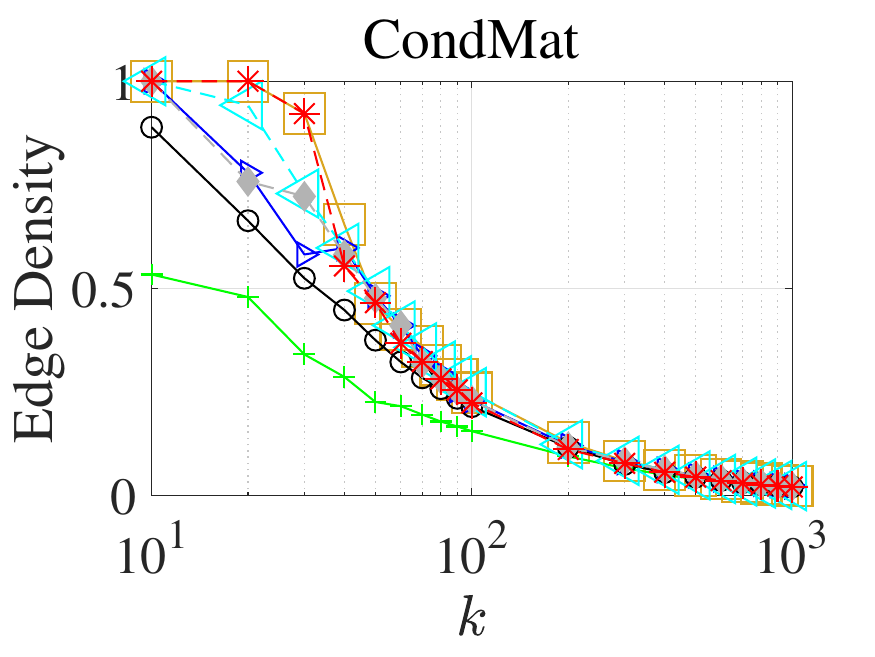}
\end{subfigure}
\vspace{0.5em}
\begin{subfigure}[b]{0.48\columnwidth}
  \centering
  \includegraphics[width=\linewidth]{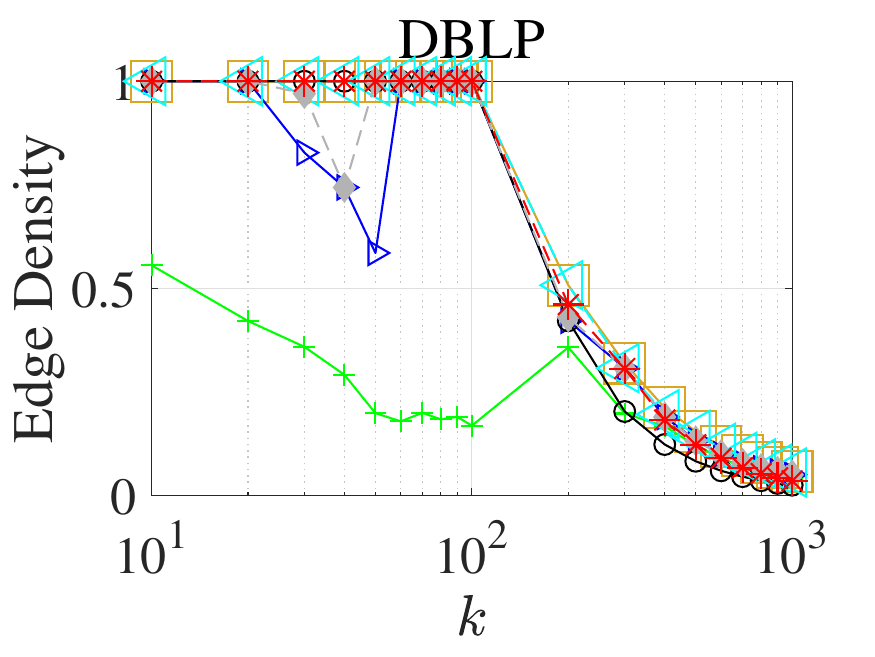}
\end{subfigure}
\hfill
\begin{subfigure}[b]{0.48\columnwidth}
  \centering
  \includegraphics[width=\linewidth]{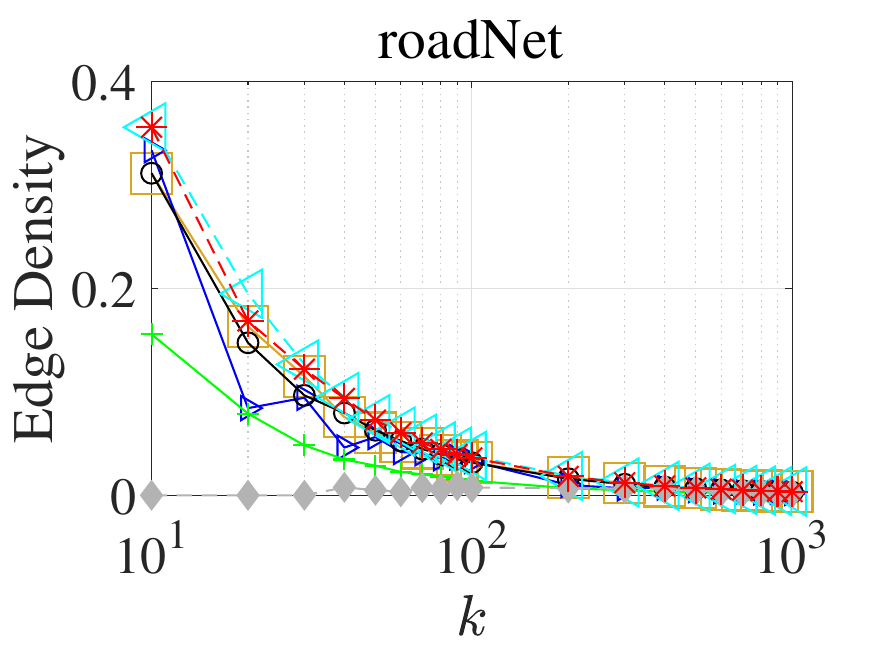}
\end{subfigure}
\vspace{0.5em}
\begin{subfigure}[b]{0.48\columnwidth}
  \centering
  \includegraphics[width=\linewidth]{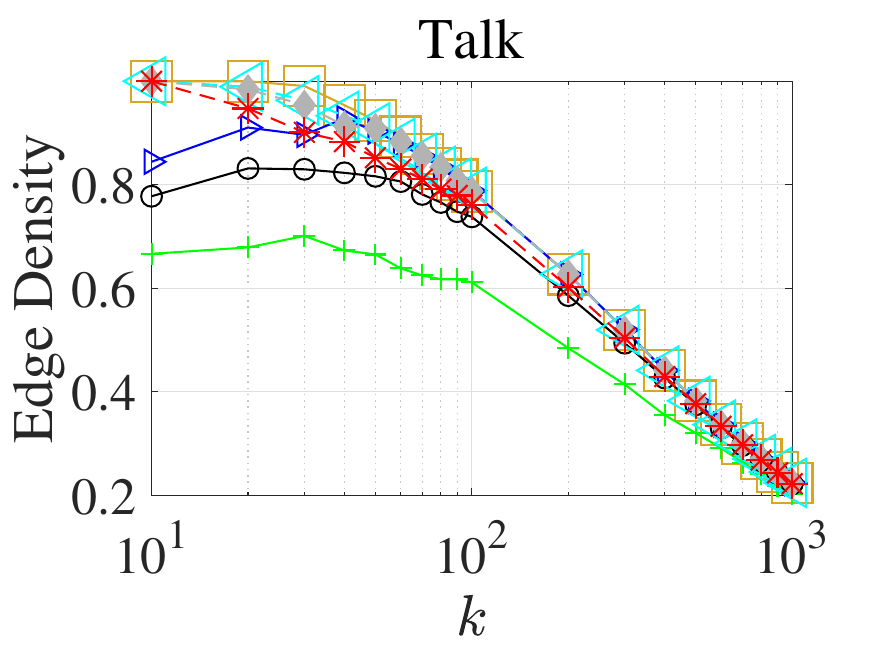}
\end{subfigure}
\hfill
\begin{subfigure}[b]{0.48\columnwidth}
  \centering
  \includegraphics[width=\linewidth]{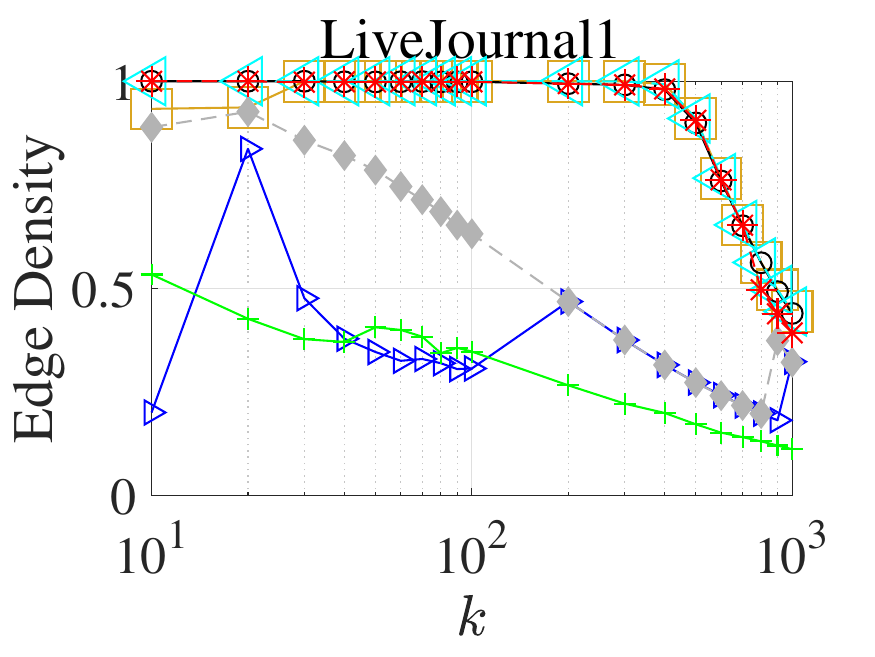}
\end{subfigure}
\begin{subfigure}[b]{0.99\columnwidth}
  \centering
  \includegraphics[width=\linewidth]{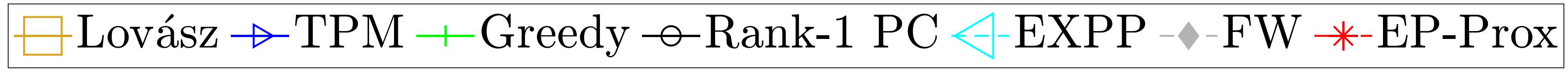}
\end{subfigure}
\caption{Edge density under different \( k \) for D$k$S problem.}
\label{fig_dks_density}
\end{figure}

\noindent \textbf{Edge Density Performance: }
Figure \ref{fig_dks_density} illustrates the density performance of different algorithms across $k$.
Our proposed EP-Prox method achieves performance competitive with that of EXPP and the Lovász relaxation on all datasets.
While Rank-1 PC, TPM, and FW can work well in some instances, they are less consistent.
Greedy, in particular, often yields significantly lower edge density.
\\\noindent \textbf{Runtime Comparison:}
Figure~\ref{fig_dks_time} reports the runtime of all the methods. Compared to the Lovász relaxation and EXPP, our EP-Prox method demonstrates significantly lower computational cost and is 2X faster across all datasets, thereby highlighting its scalability.

\begin{figure}[!htb]
\centering
\begin{subfigure}[b]{0.48\columnwidth}
  \centering
  \includegraphics[width=\linewidth]{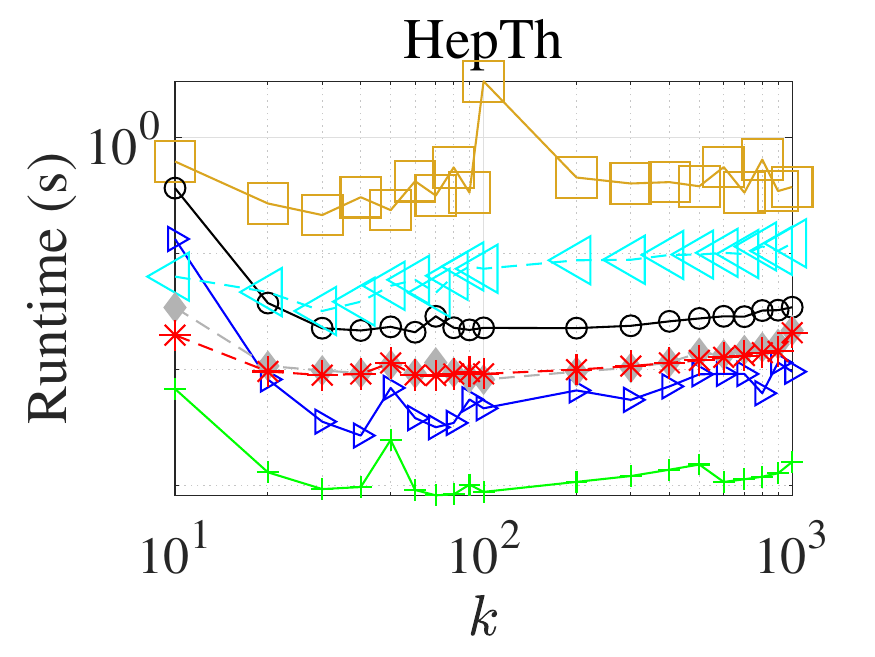}
\end{subfigure}
\hfill
\begin{subfigure}[b]{0.48\columnwidth}
  \centering
  \includegraphics[width=\linewidth]{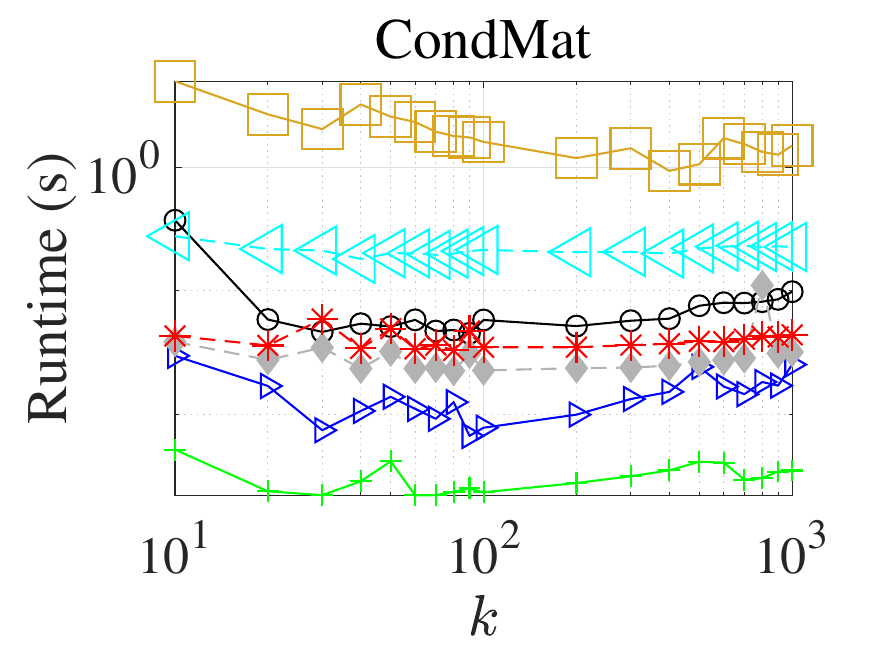}
\end{subfigure}
\vspace{0.5em}
\begin{subfigure}[b]{0.48\columnwidth}
  \centering
  \includegraphics[width=\linewidth]{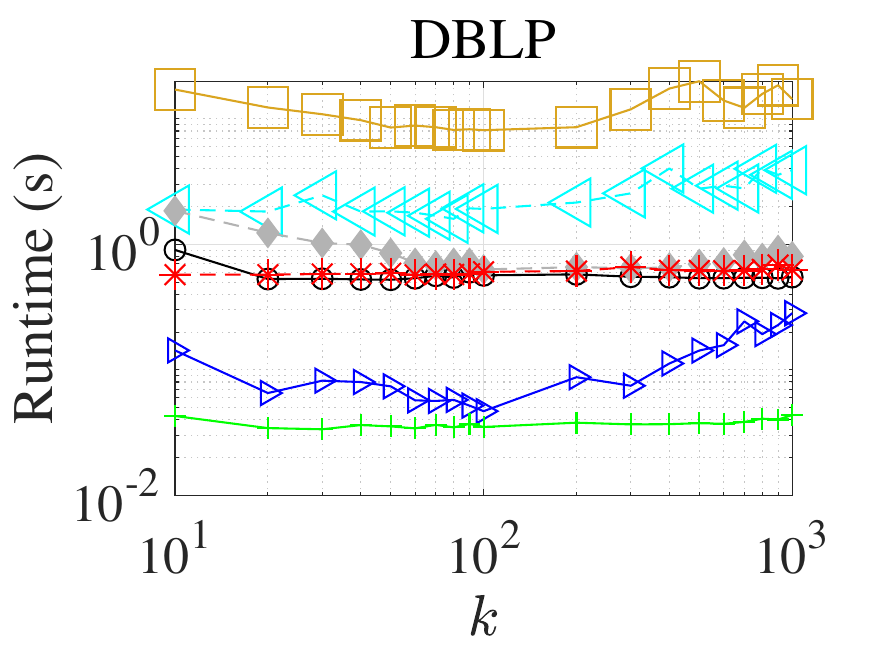}
\end{subfigure}
\hfill
\begin{subfigure}[b]{0.48\columnwidth}
  \centering
  \includegraphics[width=\linewidth]{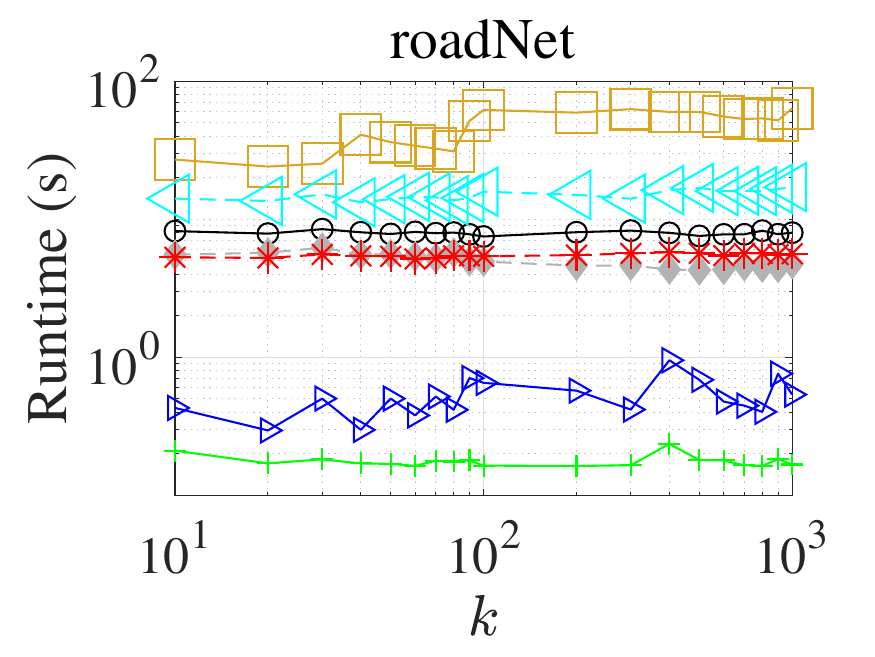}
\end{subfigure}
\vspace{0.5em}
\begin{subfigure}[b]{0.48\columnwidth}
  \centering
  \includegraphics[width=\linewidth]{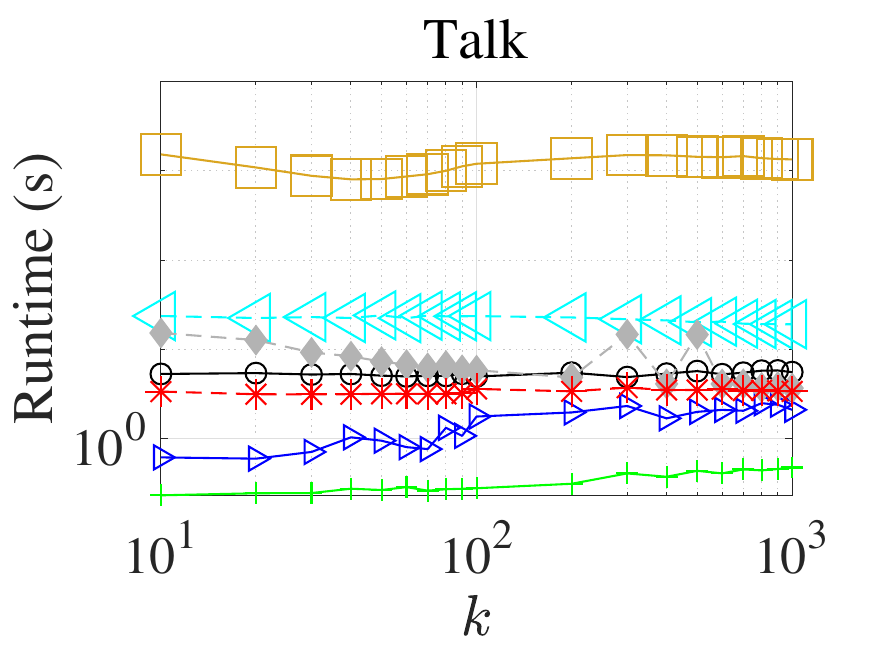}
\end{subfigure}
\hfill
\begin{subfigure}[b]{0.48\columnwidth}
  \centering
  \includegraphics[width=\linewidth]{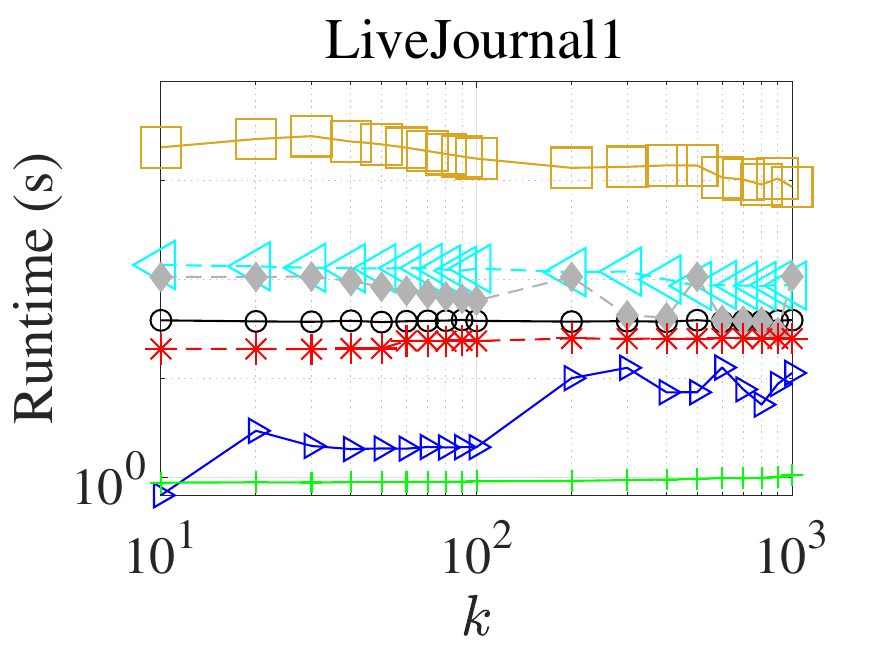}
\end{subfigure}
\begin{subfigure}[b]{0.99\columnwidth}
  \centering
  \includegraphics[width=\linewidth]{figures/dks/dks_legend.jpg}
\end{subfigure}
\caption{Runtime under different \( k \) for D$k$S problem.}
\label{fig_dks_time}
\end{figure}

\subsection{The D$k_1k_2$BS Problem}
We evaluate our method on several real-world bipartite networks obtained from the KONECT repository~\cite{kunegis2013konect}.
The selected graphs cover a wide range of sizes, and detailed statistics of the datasets used in our experiments are summarized in Table~\ref{table_stat_dkbs}.
\begin{table}[!h]
	\centering
    \setlength{\tabcolsep}{1mm}
	\begin{tabular}{c|c|c|c}
		\hline
		Datasets & $n_1$ & $n_2$& $m$ \\
		\hline\hline
		Actor movies & 127K  & 383K & 1.47M \\\hline
		Digg votes &  139K & 3,553 & 3M \\\hline
        Flickr & 396K  & 103K & 8.5M \\ \hline
		Wikipedia edits (en) & 8.1M & 42.6M & 572.6M \\ \hline
	\end{tabular}
	\caption{Statistics of the bipartite datasets, where \(n_1\), \(n_2\) denote the sizes of the two node sets, and \(m\) is the number of edges.}
	\label{table_stat_dkbs}
\end{table}

\begin{figure}[!t]
\centering
\begin{subfigure}[b]{0.48\columnwidth}
  \centering
  \includegraphics[width=\linewidth]{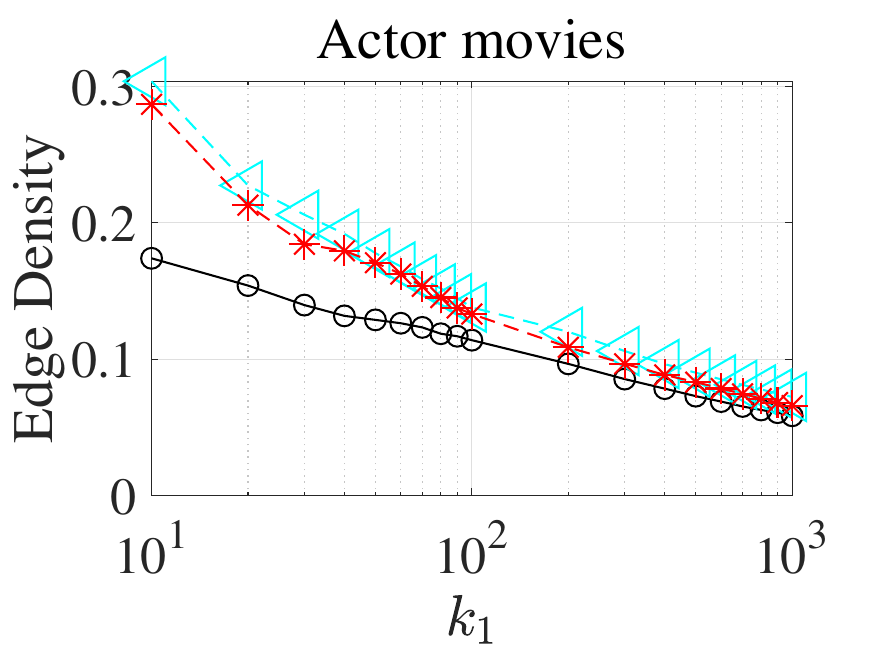}
\end{subfigure}
\hfill
\begin{subfigure}[b]{0.48\columnwidth}
  \centering
  \includegraphics[width=\linewidth]{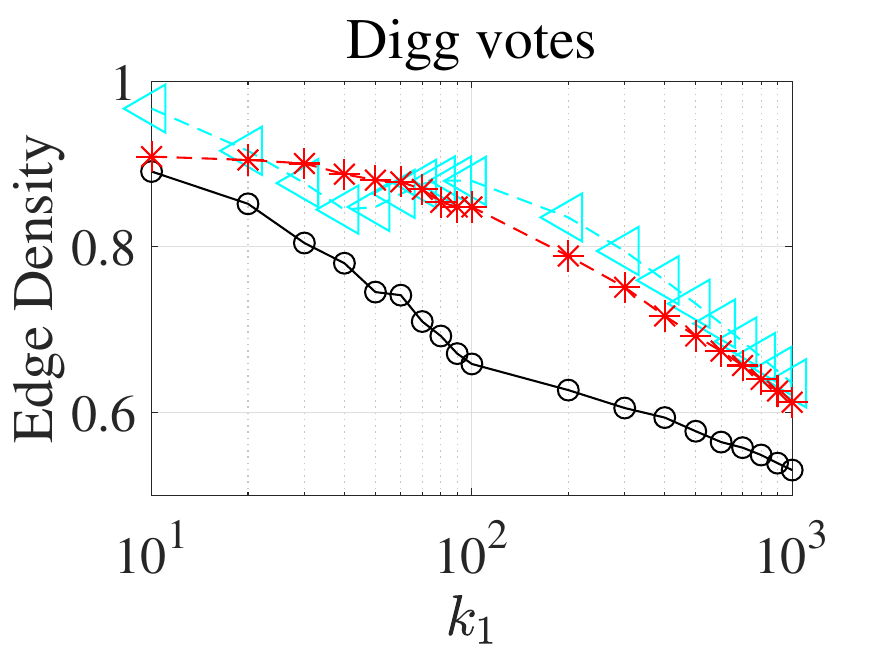}
\end{subfigure}
\vspace{0.5em}
\begin{subfigure}[b]{0.48\columnwidth}
  \centering
  \includegraphics[width=\linewidth]{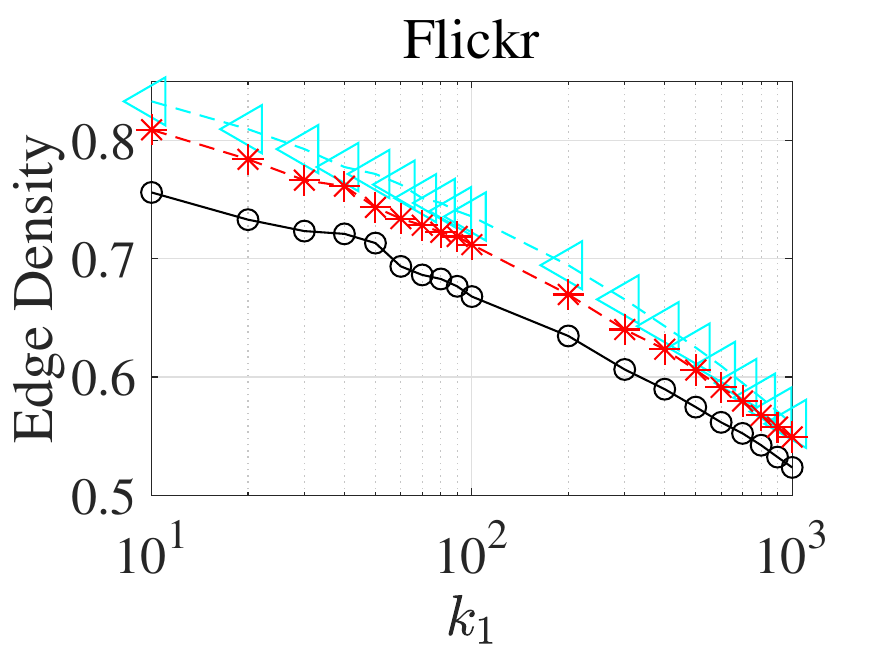}
\end{subfigure}
\hfill
\begin{subfigure}[b]{0.48\columnwidth}
  \centering
  \includegraphics[width=\linewidth]{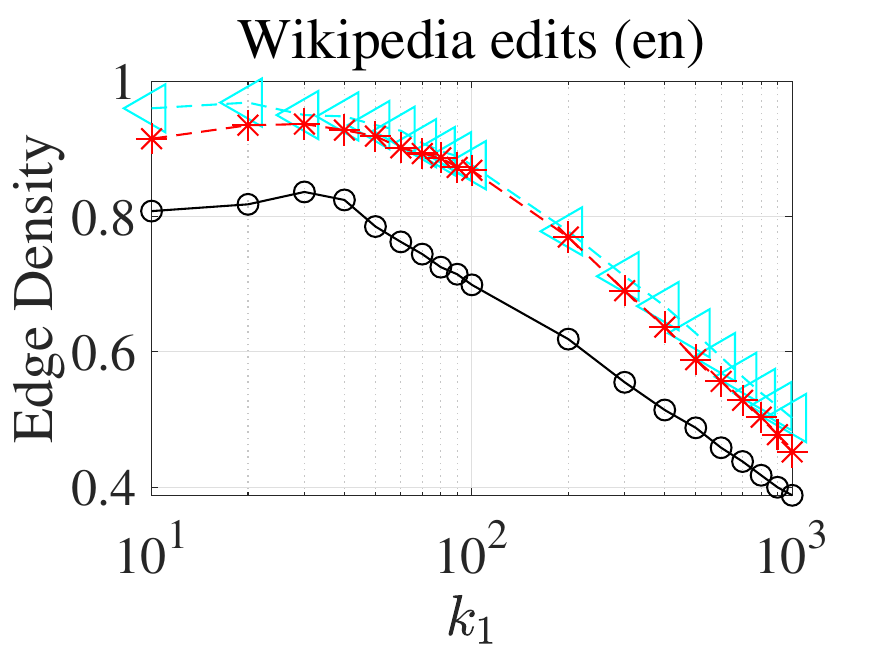}
\end{subfigure}
\begin{subfigure}[b]{0.99\columnwidth}
  \centering
  \includegraphics[width=0.45\linewidth]{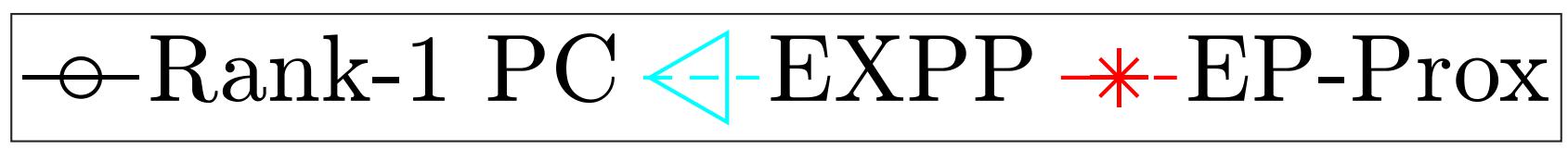}
\end{subfigure}
\caption{Edge density under different \( k_1 \) for D$k_1k_2$BS.}
\label{fig_dkbs_density_k2100}
\end{figure}

\begin{figure}[htb!]
\centering
\begin{subfigure}[b]{0.48\columnwidth}
  \centering
  \includegraphics[width=\linewidth]{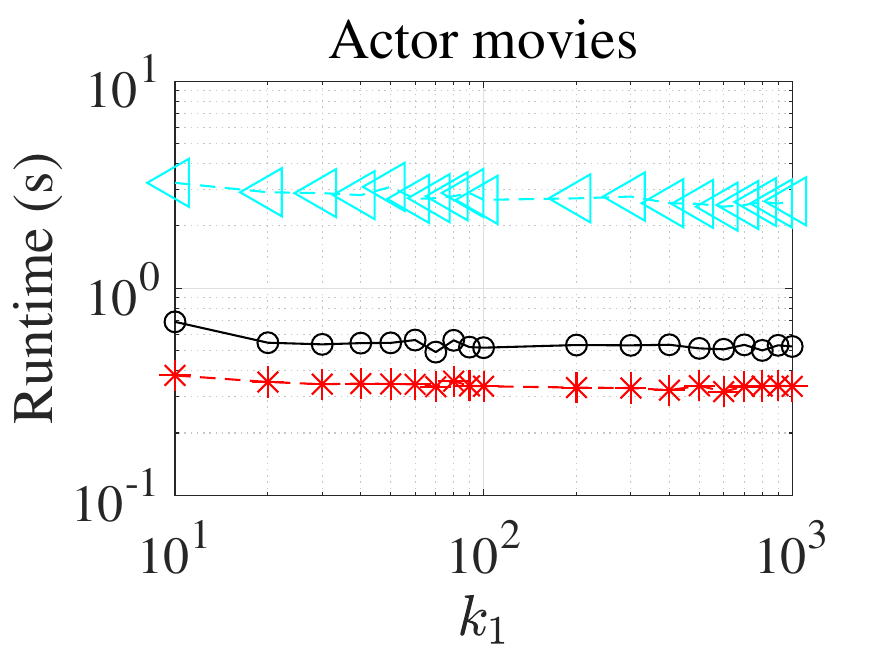}
\end{subfigure}
\hfill
\begin{subfigure}[b]{0.48\columnwidth}
  \centering
  \includegraphics[width=\linewidth]{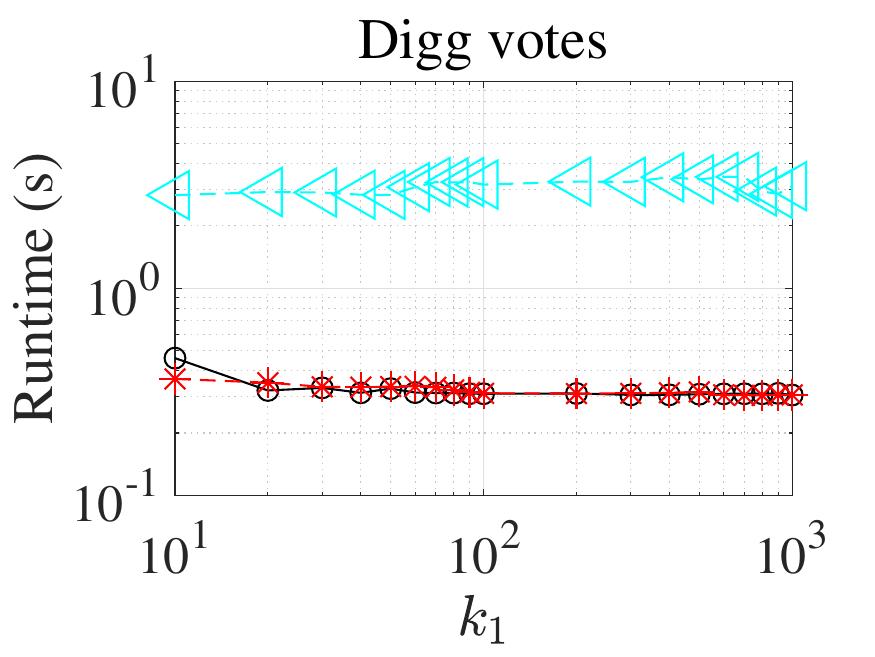}
\end{subfigure}
\vspace{0.5em}
\begin{subfigure}[b]{0.48\columnwidth}
  \centering
  \includegraphics[width=\linewidth]{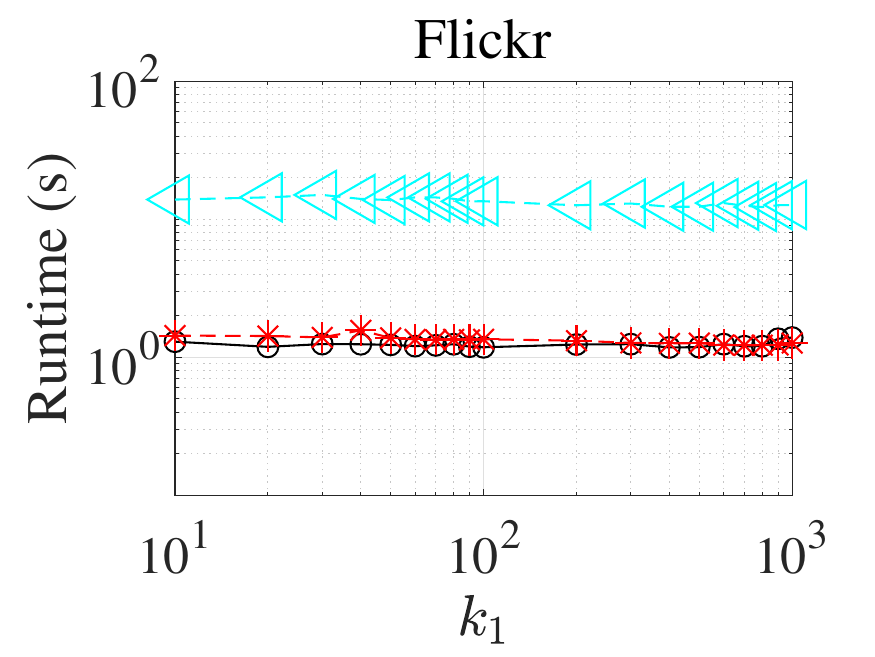}
\end{subfigure}
\hfill
\begin{subfigure}[b]{0.48\columnwidth}
  \centering
  \includegraphics[width=\linewidth]{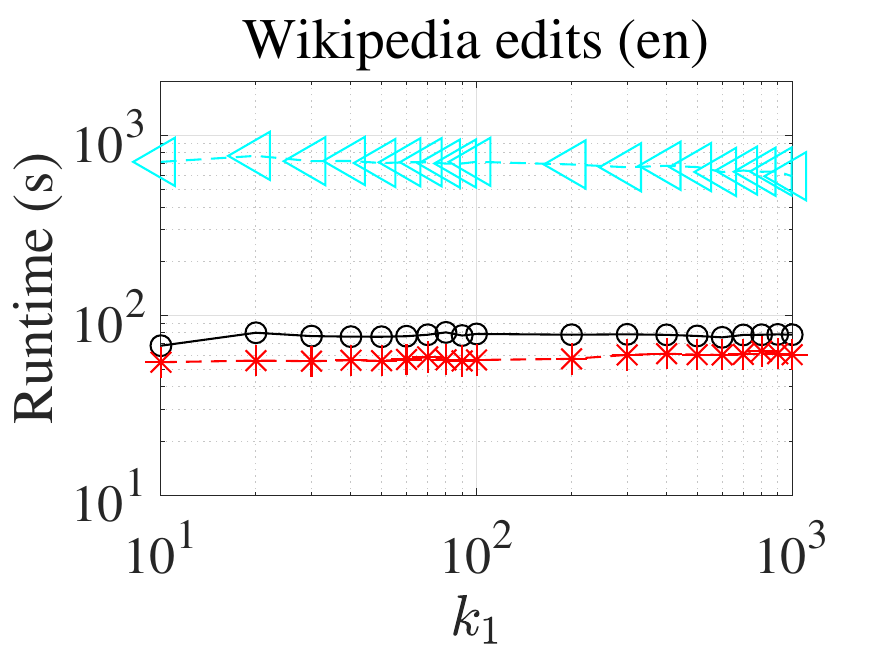}
\end{subfigure}
\begin{subfigure}[b]{0.99\columnwidth}
  \centering
  \includegraphics[width=0.45\linewidth]{figures/dkbs/dkbs_legend.jpg}
\end{subfigure}
\caption{Runtime under different \( k_1 \) for D$k_1k_2$BS.}
\label{fig_dkbs_time_k2100}
\end{figure}

To quantify the density of the extracted subgraph, we use the metric: 
\[
{\rm edge~density }=\bx^\top\bB\by/(k_1 k_2),
\]
where $\bx\in\setU^{n_1}_{k_1}$ and $\by\in\setU^{n_2}_{k_2}$ are the binary selection vectors returned by the respective methods. 
A higher edge density indicates a denser subgraph and thus better solution quality. We adapt the rank-1 PC and EXPP methods to solve the D$k_1k_2$BS problem, and use them as baselines for comparison.
This method follows the implementation in Algorithm~\ref{alg_whole}, similar to D$k$S problem.
In this test, we change the value of $k_1$ from $10$ to $1000$, and set $k_2=100$. 
The results in terms of edge density and runtime are shown in Figures~\ref{fig_dkbs_density_k2100} and \ref{fig_dkbs_time_k2100}.
Additional details on the pre-processing steps, algorithm settings, and additional results for the case of $k_1 = k_2$ are provided in Part F of the supplementary material.

 \noindent \textbf{Edge Density and Runtime Comparison:}
From Figure \ref{fig_dkbs_density_k2100}, both EXPP and EP-Prox outperform Rank-1 PC, yielding denser subgraphs. Meanwhile, from Figure \ref{fig_dkbs_time_k2100}, EP-Prox is noticeably faster than EXPP (about an order of magnitude). Hence, EP-Prox attains a better performance-complexity trade-off.

\section{Conclusion}
\label{sec_conclusion}
We proposed a novel exact penalized formulation for the D$k$S problem based on the error bound principle, preserving both global and local optima. To solve it efficiently, we developed a non-convex proximal gradient method with low per-iteration cost and provided a convergence analysis. Extensive experiments on real-world large-scale graphs demonstrate that our approach achieves a good balance between the solution quality and the computation expense. 
Future work will explore the applicability of this framework for general cardinality-constrained optimization problems.

\section{Acknowledgments}
Ya Liu, Junbin Liu, and Wing-Kin Ma were supported by a General Research Fund (GRF) of Hong Kong Research Grant Council (RGC) under Project ID CUHK 14203721 and by a CUHK Direct Grant under Project ID 4055268. 
Aritra Konar was supported by KU Leuven Special Research Fund BOF/STG-22-040.

\bibliography{ref.bib}

@misc{snapnets,
author       = {Jure Leskovec and Andrej Krevl},
title        = {{SNAP Datasets}: {Stanford} Large Network Dataset Collection},
howpublished = {\url{http://snap.stanford.edu/data}},
month        = jun,
year         = 2014
}

@article{liu2024extreme,
  title={Extreme Point Pursuit—Part {I}: A Framework for Constant Modulus Optimization},
  author={Liu, Junbin and Liu, Ya and Ma, Wing-Kin and Shao, Mingjie and So, Anthony Man-Cho},
  journal={IEEE Transactions on Signal Processing},
  year={2024},
  publisher={IEEE}
}

@article{liu2024extremeII,
  title={Extreme Point Pursuit—Part {II}: Further Error Bound Analysis and Applications},
  author={Liu, Junbin and Liu, Ya and Ma, Wing-Kin and Shao, Mingjie and So, Anthony Man-Cho},
  journal={IEEE Transactions on Signal Processing},
  year={2024},
  publisher={IEEE}
}

@book{luo1996mathematical,
  title={Mathematical {P}rograms with {E}quilibrium {C}onstraints},
  author={Luo, Zhi-Quan and Pang, Jong-Shi and Ralph, Daniel},
  year={1996},
  publisher={Cambridge University Press}
}

@inproceedings{lu2025densest,
  title={Densest $k$-subgraph mining via a provably tight relaxation},
  author={Lu, Qiheng and Sidiropoulos, Nicholas D and Konar, Aritra},
  booktitle={Proceedings of the AAAI Conference on Artificial Intelligence},
  volume={39},
  pages={12291--12299},
  year={2025}
}

@article{motzkin1965maxima,
  title={Maxima for graphs and a new proof of a theorem of Tur{\'a}n},
  author={Motzkin, Theodore S and Straus, Ernst G},
  journal={Canadian Journal of Mathematics},
  volume={17},
  pages={533--540},
  year={1965},
  publisher={Cambridge University Press}
}

@book{nocedal2006numerical,
  title={Numerical {O}ptimization},
  author={Nocedal, Jorge and Wright, Stephen J},
  year={2006},
  publisher={Springer}
}

@article{keys2019proximal,
  title={Proximal distance algorithms: Theory and practice},
  author={Keys, Kevin L and Zhou, Hua and Lange, Kenneth},
  journal={Journal of Machine Learning Research},
  volume={20},
  number={66},
  pages={1--38},
  year={2019}
}

@inproceedings{liu2024cardinality,
  title={Cardinality-constrained binary quadratic optimization via extreme point pursuit, with application to the densest $k$-subgraph problem},
  author={Liu, Ya and Liu, Junbin and Ma, Wing-Kin},
  booktitle={ICASSP 2024-2024 IEEE International Conference on Acoustics, Speech and Signal Processing (ICASSP)},
  pages={9631--9635},
  year={2024},
  organization={IEEE}
}

@article{zaslavskiy2008path,
  title={A path following algorithm for the graph matching problem},
  author={Zaslavskiy, Mikhail and Bach, Francis and Vert, Jean-Philippe},
  journal={IEEE {T}ransactions on {P}attern {A}nalysis and {M}achine {I}ntelligence},
  volume={31},
  number={12},
  pages={2227--2242},
  year={2008},
  publisher={IEEE}
}

@article{shao2020binary,
  title={Binary {MIMO} detection via homotopy optimization and its deep adaptation},
  author={Shao, Mingjie and Ma, Wing-Kin},
  journal={IEEE Transactions on Signal Processing},
  volume={69},
  pages={781--796},
  year={2020},
  publisher={IEEE}
}

@article{xu2017globally,
  title={A globally convergent algorithm for nonconvex optimization based on block coordinate update},
  author={Xu, Yangyang and Yin, Wotao},
  journal={Journal of Scientific Computing},
  volume={72},
  number={2},
  pages={700--734},
  year={2017},
  publisher={Springer}
}

@book{rockafellar2009variational,
	title={Variational {A}nalysis},
	author={Rockafellar, R Tyrrell and Wets, Roger J-B},
	volume={317},
	year={2009},
	publisher={Springer}
}

@article{li2020understanding,
  title={Understanding notions of stationarity in nonsmooth optimization: A guided tour of various constructions of subdifferential for nonsmooth functions},
  author={Li, Jiajin and So, Anthony Man-Cho and Ma, Wing-Kin},
  journal={IEEE Signal Processing Magazine},
  volume={37},
  number={5},
  pages={18--31},
  year={2020},
  publisher={IEEE}
}

@inproceedings{kunegis2013konect,
  title={Konect: the koblenz network collection},
  author={Kunegis, J{\'e}r{\^o}me},
  booktitle={Proceedings of the 22nd {I}nternational {C}onference on {W}orld {W}ide {W}eb},
  pages={1343--1350},
  year={2013}
}

@article{yuan2013truncated,
  title={Truncated power method for sparse eigenvalue problems},
  author={Yuan, Xiao-Tong and Zhang, Tong},
  journal={The Journal of Machine Learning Research},
  volume={14},
  number={1},
  pages={899--925},
  year={2013},
  publisher={JMLR. org}
}

@book{feige1997densest,
  title={On the {D}ensest $k$-{S}ubgraph {P}roblem},
  author={Feige, Uriel and Seltser, Michael and others},
  year={1997},
  publisher={Weizmann Institute of Science.}
}

@article{karisch2000solving,
  title={Solving graph bisection problems with semidefinite programming},
  author={Karisch, Stefan E and Rendl, Franz and Clausen, Jens},
  journal={INFORMS Journal on Computing},
  volume={12},
  number={3},
  pages={177--191},
  year={2000},
  publisher={INFORMS}
}

@book{CP22,
  title={Modern {N}onconvex {N}ondifferentiable {O}ptimization},
  author={Cui, Ying and Pang, Jong-Shi},
  year={2021},
  publisher={SIAM}
}

@book{garey2002computers,
  title={Computers and intractability},
  author={Garey, Michael R and Johnson, David S},
  volume={29},
  year={2002},
  publisher={wh freeman New York}
}

@inproceedings{shin2016corescope,
  title={Corescope: {G}raph mining using k-core analysis—patterns, anomalies and algorithms},
  author={Shin, Kijung and Eliassi-Rad, Tina and Faloutsos, Christos},
  booktitle={IEEE 16th {I}nternational {C}onference on {D}ata {M}ining (ICDM)},
  pages={469--478},
  year={2016},
  organization={IEEE}
}

@book{nesterov2018lectures,
  title={Lectures on {C}onvex {O}ptimization},
  author={Nesterov, Yurii},
  volume={137},
  year={2018},
  publisher={Springer}
}

@inproceedings{harb2023convergence,
  title={Convergence to Lexicographically Optimal Base in a (Contra) Polymatroid and Applications to Densest Subgraph and Tree Packing},
  author={Harb, Elfarouk and Quanrud, Kent and Chekuri, Chandra},
  booktitle={31st Annual European Symposium on Algorithms (ESA 2023)},
  volume={274},
  pages={56},
  year={2023},
  organization={Schloss Dagstuhl--Leibniz-Zentrum f{\"u}r Informatik}
}

@article{sotirov2020solving,
title={On Solving the Densest $k$-Subgraph Problem on Large Graphs},
author={Sotirov, Renata},
journal={Optimization Methods and Software},
volume={35},
number={6},
pages={1160-1178},
year={2020}
}

@article{hager2016projection,
title={Projection Algorithms for Nonconvex Minimization with Application to Sparse Principal Component Analysis},
author={Hager, William W and Phan, Dzung T and Zhu, Jiajie},
journal={Journal of Global Optimization},
volume={65},
pages={657-676},
year={2016}
}

@inproceedings{bombina2020convex,
title={Convex Optimization for the Densest Subgraph and Densest Submatrix Problems},
author={Bombina, Polina and Ames, Brendan},
booktitle={SN Operations Research Forum},
volume={1},
pages={1-24},
year={2020},
publisher={Springer}
}

@inproceedings{bhaskara2010detecting,
title={Detecting High Log-Densities: an $O (n^{1/4})$ Approximation for Densest $k$-Subgraph},
author={Bhaskara, Aditya and Charikar, Moses and Chlamtac, Eden and Feige, Uriel and Vijayaraghavan, Aravindan},
booktitle={Proceedings of the Forty-Second ACM Symposium on Theory of Computing},
pages={201-210},
year={2010},
publisher={ACM}
}

@article{kawase2018densest,
  title={The densest subgraph problem with a convex/concave size function},
  author={Kawase, Yasushi and Miyauchi, Atsushi},
  journal={Algorithmica},
  volume={80},
  number={12},
  pages={3461--3480},
  year={2018},
  publisher={Springer}
}

@article{beck2009fast,
  title={A fast iterative shrinkage-thresholding algorithm for linear inverse problems},
  author={Beck, Amir and Teboulle, Marc},
  journal={SIAM {J}ournal on {I}maging {S}ciences},
  volume={2},
  number={1},
  pages={183--202},
  year={2009},
  publisher={SIAM}
}

@article{harb2022faster,
  title={Faster and scalable algorithms for densest subgraph and decomposition},
  author={Harb, Elfarouk and Quanrud, Kent and Chekuri, Chandra},
  journal={Advances in Neural Information Processing Systems},
  volume={35},
  pages={26966--26979},
  year={2022}
}

@inproceedings{nguyen2024multiplicative,
  title={Multiplicative Weights Update, Area Convexity and Random Coordinate Descent for Densest Subgraph Problems},
  author={Nguyen, Ta Duy and Ene, Alina},
  booktitle={International Conference on Machine Learning},
  pages={37683--37706},
  year={2024},
  organization={PMLR}
}

@inproceedings{danisch2017large,
  title={Large scale density-friendly graph decomposition via convex programming},
  author={Danisch, Maximilien and Chan, T-H Hubert and Sozio, Mauro},
  booktitle={Proceedings of the 26th International Conference on World Wide Web},
  pages={233--242},
  year={2017}
}

@inproceedings{veldt2021generalized,
  title={The generalized mean densest subgraph problem},
  author={Veldt, Nate and Benson, Austin R and Kleinberg, Jon},
  booktitle={Proceedings of the 27th ACM SIGKDD Conference on Knowledge Discovery and Data Mining},
  pages={1604--1614},
  year={2021}
}

@article{seidman1983network,
  title={Network structure and minimum degree},
  author={Seidman, Stephen B},
  journal={Social {N}etworks},
  volume={5},
  number={3},
  pages={269--287},
  year={1983},
  publisher={Elsevier}
}

@inproceedings{andersen2009finding,
  title={Finding dense subgraphs with size bounds},
  author={Andersen, Reid and Chellapilla, Kumar},
  booktitle={International {W}orkshop on {A}lgorithms and {M}odels for the {W}eb-{G}raph},
  pages={25--37},
  year={2009},
  organization={Springer}
}

@inproceedings{konar2021exploring,
  title={Exploring the subgraph density-size trade-off via the Lova{\'s}z extension},
  author={Konar, Aritra and Sidiropoulos, Nicholas D},
  booktitle={Proceedings of the 14th ACM International Conference on Web Search and Data Mining},
  pages={743--751},
  year={2021}
}

@inproceedings{papailiopoulos2014finding,
  title={Finding dense subgraphs via low-rank bilinear optimization},
  author={Papailiopoulos, Dimitris and Mitliagkas, Ioannis and Dimakis, Alexandros and Caramanis, Constantine},
  booktitle={International Conference on Machine Learning},
  pages={1890--1898},
  year={2014},
  organization={PMLR}
}

@inproceedings{charikar2000greedy,
  title={Greedy approximation algorithms for finding dense components in a graph},
  author={Charikar, Moses},
  booktitle={International Workshop on Approximation Algorithms for Combinatorial Optimization},
  pages={84--95},
  year={2000},
  organization={Springer}
}

@article{asahiro2000greedily,
  title={Greedily finding a dense subgraph},
  author={Asahiro, Yuichi and Iwama, Kazuo and Tamaki, Hisao and Tokuyama, Takeshi},
  journal={Journal of Algorithms},
  volume={34},
  number={2},
  pages={203--221},
  year={2000},
  publisher={Elsevier}
}

@book{goldberg1984finding,
  title={Finding a {M}aximum {D}ensity {S}ubgraph},
  author={Goldberg, Andrew V},
  year={1984},
  publisher={Technical report, University of California Berkeley, CA}
}

@article{feige2001dense,
  title={The dense k-subgraph problem},
  author={Feige, Uriel and Peleg, David and Kortsarz, Guy},
  journal={Algorithmica},
  volume={29},
  number={3},
  pages={410--421},
  year={2001},
  publisher={Springer}
}

@inproceedings{bhaskara2012polynomial,
  title={Polynomial integrality gaps for strong sdp relaxations of densest k-subgraph},
  author={Bhaskara, Aditya and Charikar, Moses and Guruswami, Venkatesan and Vijayaraghavan, Aravindan and Zhou, Yuan},
  booktitle={Proceedings of the {T}wenty-{T}hird {A}nnual ACM-SIAM {S}ymposium on Discrete Algorithms},
  pages={388--405},
  year={2012},
  organization={SIAM}
}

@inproceedings{manurangsi2017almost,
  title={Almost-polynomial ratio ETH-hardness of approximating densest k-subgraph},
  author={Manurangsi, Pasin},
  booktitle={Proceedings of the 49th Annual ACM SIGACT Symposium on Theory of Computing},
  pages={954--961},
  year={2017}
}

@inproceedings{chekuri2022densest,
  title={Densest Subgraph: Supermodularity, Iterative Peeling, and Flow},
  author={Chekuri, Chandra and Quanrud, Kent and Torres, Manuel R},
  booktitle={Proceedings of the 2022 Annual ACM-SIAM Symposium on Discrete Algorithms (SODA)},
  pages={1531--1555},
  year={2022},
  organization={SIAM}
}

@inproceedings{boob2020flowless,
  title={Flowless: Extracting densest subgraphs without flow computations},
  author={Boob, Digvijay and Gao, Yu and Peng, Richard and Sawlani, Saurabh and Tsourakakis, Charalampos and Wang, Di and Wang, Junxing},
  booktitle={Proceedings of The Web Conference 2020},
  pages={573--583},
  year={2020}
}

@article{lanciano2024survey,
  title={A survey on the densest subgraph problem and its variants},
  author={Lanciano, Tommaso and Miyauchi, Atsushi and Fazzone, Adriano and Bonchi, Francesco},
  journal={ACM Computing Surveys},
  volume={56},
  number={8},
  pages={1--40},
  year={2024},
  publisher={ACM New York, NY}
}

@misc{luo2023survey,
  title        = {A Survey of Densest Subgraph Discovery on Large Graphs},
  author       = {Luo, Wensheng and Ma, Chenhao and Fang, Yixiang and Lakshman, Laks VS},
  year         = {2023},
  eprint       = {2306.07927},
  archivePrefix= {arXiv},
  primaryClass = {cs.SI},
  note         = {arXiv preprint arXiv:2306.07927}
}

@inproceedings{khuller2009finding,
  title={On finding dense subgraphs},
  author={Khuller, Samir and Saha, Barna},
  booktitle={International {C}olloquium on {A}utomata, {L}anguages, and {P}rogramming},
  pages={597--608},
  year={2009},
  organization={Springer}
}

@inproceedings{saha2010dense,
  title={Dense subgraphs with restrictions and applications to gene annotation graphs},
  author={Saha, Barna and Hoch, Allison and Khuller, Samir and Raschid, Louiqa and Zhang, Xiao-Ning},
  booktitle={Research in Computational Molecular Biology: 14th Annual International Conference, RECOMB 2010, Lisbon, Portugal, April 25-28, 2010. Proceedings 14},
  pages={456--472},
  year={2010},
  organization={Springer}
}

@inproceedings{chen2022antibenford,
  title={Antibenford subgraphs: Unsupervised anomaly detection in financial networks},
  author={Chen, Tianyi and Tsourakakis, Charalampos},
  booktitle={Proceedings of the 28th ACM SIGKDD Conference on Knowledge Discovery and Data Mining},
  pages={2762--2770},
  year={2022}
}

@inproceedings{hooi2016fraudar,
  title={Fraudar: Bounding graph fraud in the face of camouflage},
  author={Hooi, Bryan and Song, Hyun Ah and Beutel, Alex and Shah, Neil and Shin, Kijung and Faloutsos, Christos},
  booktitle={Proceedings of the 22nd ACM SIGKDD {I}nternational {C}onference on {K}nowledge {D}iscovery and {D}ata {M}ining},
  pages={895--904},
  year={2016}
}

@inproceedings{tsourakakis2013denser,
  title={Denser than the densest subgraph: extracting optimal quasi-cliques with quality guarantees},
  author={Tsourakakis, Charalampos and Bonchi, Francesco and Gionis, Aristides and Gullo, Francesco and Tsiarli, Maria},
  booktitle={Proceedings of the 19th ACM SIGKDD {I}nternational {C}onference on {K}nowledge {D}iscovery and {D}ata {M}ining},
  pages={104--112},
  year={2013}
}

@article{angel2014dense,
  title={Dense subgraph maintenance under streaming edge weight updates for real-time story identification},
  author={Angel, Albert and Koudas, Nick and Sarkas, Nikos and Srivastava, Divesh and Svendsen, Michael and Tirthapura, Srikanta},
  journal={The VLDB {J}ournal},
  volume={23},
  number={2},
  pages={175--199},
  year={2014},
  publisher={Springer}
}

@inproceedings{li2020flowscope,
  title={Flowscope: Spotting money laundering based on graphs},
  author={Li, Xiangfeng and Liu, Shenghua and Li, Zifeng and Han, Xiaotian and Shi, Chuan and Hooi, Bryan and Huang, He and Cheng, Xueqi},
  booktitle={Proceedings of the AAAI {C}onference on {A}rtificial aintelligence},
  volume={34},
  pages={4731--4738},
  year={2020}
}

\clearpage
\newpage
\appendix

\section{A: Proof of Proposition \ref{prop:ep}}
\label{proof_prop1}
The equivalence in terms of the global optimal solution sets between problem (\ref{eq_dks}) and its penalized reformulation \eqref{eq_dks_01} follows directly from Lemma~\ref{lem:ep2}.

We prove the equivalence of local optimal solution sets. 
To begin with, we revisit an existing result stated in Lemma~\ref{lem:locmin_eb}, where we use the notation ${\rm arglocmin}_{\bx \in \setX} \, \phi(\bx)$ to denote the set of locally optimal solutions to the problem $\min_{\bx \in \setX} \phi(\bx)$.

\begin{Lemma}{\bf (Proposition 9.1.2 in \cite{CP22})} \label{lem:locmin_eb}
Consider the same setting in Lemma \ref{lem:ep2}.
Then:
\begin{itemize}
\item Given any scalar $\lambda > K$, it holds that
\[
\check{\bx} \in \mathop{\mathrm{arglocmin}}_{\bx \in \setV} \, \phi(\bx)
\Rightarrow
\check{\bx} \in \mathop{\mathrm{arglocmin}}_{\bx \in \setX} \, \phi(\bx) + \lambda \psi(\bx).
\]
\item If $\check{\bx}$ is a point in $\setV$, then the following implication is true
\[
\check{\bx} \in \mathop{\mathrm{arglocmin}}_{\bx \in \setX} \, \phi(\bx) + \lambda \psi(\bx)
\Rightarrow
\check{\bx} \in \mathop{\mathrm{arglocmin}}_{\bx \in \setV} \, \phi(\bx).
\]
\end{itemize}
\end{Lemma}
\medskip
This result needs the assumption $\check{\bx}\in\setV$ to guarantee the equivalence of locally optimal solutions. 
We show that our proposed penalization formulation eliminates this assumption, leading to equivalence between locally optimal solutions.
We begin the proof. 
Let $\bx\in[0,1]^n$.
By Lemma~\ref{thm_eb2_Unk_01}, the distance from $\bx$ to the set $\mathcal{U}^n_\kappa$ satisfies:
\begin{equation}
\mathrm{dist}(\bx, \mathcal{U}^n_k)
 \leq k + \underbrace{\bone^\top \bx -2S_k(\bx)}_{=h(\bx)}.
\end{equation}
Now, suppose $ \check{\bx}\in \mathop{\rm arglocmin}_{\bx\in[0,1]^n}{F}_\lambda(\bx) := {f}(\bx) + \lambda h(\bx) $. 
Without loss of generality, assume $\check{x}_1\geq \dots\geq \check{x}_n$.
By the definition of local optimality, there exists a scalar $\epsilon > 0$ such that:
\begin{equation}
{F}_\lambda(\check{\bx}) \leq {F}_\lambda(\bx), \forall \bx\in [0,1]^n, \Arrowvert \bx-\check{\bx}\Arrowvert_2\leq\epsilon.
\label{eq_star}
\end{equation}
Let $\bx$ be defined as follows:
\[
x_i =
\left\{
\begin{array}{ll}
\check{x}_i, & \mathrm{if } ~ i = 1, \dots, k, \, k+2, \dots, n \\
\max\{0, \check{x}_{k+1} - \epsilon\}, & \mathrm{if }~ i = k+1.
\end{array}
\right.
\]
Note that if $\check{\bx} \notin \mathcal{U}^n_k$, then we must have $\check{x}_{k+1} > 0$. It follows that $\bx \neq \check{\bx}$, $\bx \in [0,1]^n$, and $\Arrowvert\bx - \check{\bx}\Arrowvert_2 \leq \epsilon$. Furthermore, we have:
\begin{equation}
\begin{array}{rl}
F_\lambda(\check{\bx}) \geq & f(\bx) - K \|\bx - \check{\bx}\|_2 + \lambda h(\check{\bx}) \\[4pt]
= & F_\lambda(\bx) - K \|\bx - \check{\bx}\|_2 + \lambda \big( h(\check{\bx}) - h(\bx) \big) \\[4pt]
> & F_\lambda(\bx),
\end{array}
\label{eq_condict}
\end{equation}
where for the last inequality,  we have $ \lambda > K $ and
\begin{equation}
\begin{array}{rl}
h(\check{\bx}) - h(\bx) 
= & \check{x}_{k+1} - x_{k+1} \\[4pt]
= & | \check{x}_{k+1} - x_{k+1} | \\[4pt]
= & \| \check{\bx} - \bx \|_2.
\end{array}
\label{eq_diff_h}
\end{equation}
The result in equation~(\ref{eq_condict}) contradicts~(\ref{eq_star}).

Thus, for $\lambda > K$, $\check{\bx}$ must satisfy $\check{\bx} \in \mathcal{U}^n_k$. 
Consequently, we can conclude that:
\begin{equation}
\mathop{\rm arglocmin}_{\bx\in\mathcal{U}^n_k} {f}(\bx) =
\mathop{\rm arglocmin}_{\bx\in[0,1]^n } {f}(\bx) + \lambda h(\bx).
\end{equation}
This establishes the equivalence of the local optimality of problem (\ref{eq_dks}) and its penalized reformulation \eqref{eq_dks_01}.
\hfill $\blacksquare$
\medskip

\section{B: Proof of Proposition~\ref{lem_dks_prox}}
\label{proof_lem_prox}
The non-convex optimization problem can be rewritten as:
\begin{equation}
\label{eq:p}
\min_{\bx \in [0,1]^n} u(\bx) = \frac{1}{2} \| \bx - \bz \|_2^2 + \mu \left( \sum_{i=k+1}^n x_{[i]} - \sum_{i=1}^k x_{[i]} \right),
\end{equation}
where \( k \leq n \) is an integer, and \( x_{[i]} \) denotes the \( i \)-th largest element of \( \bx \). Without loss of generality, assume \( z_1 \geq z_2 \geq \cdots \geq z_n \).
The objective can be rewritten as:
\begin{equation}
\label{eq:ux}
\begin{array}{l}
u(\bx) = \sum_{i=1}^k \left( \frac{1}{2} (x_{[i]} - z_{j_i(\bx)})^2 - \mu x_{[i]} \right) \\
\quad + \sum_{i=k+1}^n \left( \frac{1}{2} (x_{[i]} - z_{j_i(\bx)})^2 + \mu x_{[i]} \right),
\end{array}
\end{equation}
where \( j_i(\bx) \) is the index of \( x_{[i]} \) in \( \bx \), which depends on \( \bx \).
It can be verified that when
\begin{equation}
x_{[i]} =
\left\{
\begin{array}{ll}
\left[ z_{j_i(\bx)} + \mu \right]_0^1, & \mathrm{if~} i \leq k, \\
\left[ z_{j_i(\bx)} - \mu \right]_0^1, & \mathrm{if~} i > k,
\end{array}
\right.
\end{equation}
we have
\begin{equation}
u(\bx) \geq \tilde{u}(\bx),
\end{equation}
where
\begin{equation}
\label{eq:utilde}
\begin{array}{l}
\tilde{u}(\bx) 
\\= \sum_{i=1}^k \left( \frac{1}{2} \left( [z_{j_i(\bx)} + \mu]_0^1 - z_{j_i(\bx)} \right)^2 - \mu [z_{j_i(\bx)} + \mu]_0^1 \right) \\
+ \sum_{i=k+1}^n \left( \frac{1}{2} \left( [z_{j_i(\bx)} - \mu]_0^1 - z_{j_i(\bx)} \right)^2 + \mu [z_{j_i(\bx)} - \mu]_0^1 \right).
\end{array}
\end{equation}
Note that \( \tilde{u} \) is still a function of \( \bx \). 
To proceed, we need the following lemma.

\begin{Lemma}
\label{lem2}
For a given \( \mu > 0 \), define the function
\begin{equation}
\label{eq:g}
\begin{array}{l}
g(z) = \left( \frac{1}{2} (z - [z + \mu]_0^1)^2 - \mu [z + \mu]_0^1 \right) \\
\quad - \left( \frac{1}{2} (z - [z - \mu]_0^1)^2 + \mu [z - \mu]_0^1 \right).
\end{array}
\end{equation}
Then \( g(z) \) is monotonically non-increasing.
\end{Lemma}

\medskip
{\em Proof of Lemma \ref{lem2}:}
 We can evaluate the following functions segment-wise:
 \begin{equation}
[z+\mu]_0^1 =
\left\{
\begin{array}{ll}
1, & \text{if } z\geq 1 - \mu, \\
z + \mu, & \text{if } -\mu \leq z < 1 - \mu, \\
0, & \text{if } z< -\mu,
\end{array}
\right.
\end{equation}
\begin{equation}
[z - \mu]_0^1 =
\left\{
\begin{array}{ll}
1, & \text{if } z \geq 1 + \mu, \\
z - \mu, & \text{if } \mu \leq z < 1 + \mu, \\
0, & \text{if } z< \mu,
\end{array}
\right.
\end{equation}

\begin{equation}
(z - [z+\mu]_0^1)^2 =
\left\{
\begin{array}{ll}
(z - 1)^2, & \mathrm{if } ~ z\geq 1 - \mu, \\
\mu^2, & \mathrm{if } ~ -\mu \leq z < 1 - \mu, \\
z^2, & \mathrm{if } ~z < -\mu,
\end{array}
\right.
\end{equation}

\begin{equation}
(z - [z - \mu]_0^1)^2 =
\left\{
\begin{array}{ll}
(z - 1)^2, & \mathrm{if }~ z \geq 1 + \mu, \\
\mu^2, & \mathrm{if }~ \mu \leq z < 1 + \mu, \\
z^2, & \mathrm{if }~ z < \mu.
\end{array}
\right.
\end{equation}
Then we are ready to write out the expression for \( g(z) \).  
There are two cases, \( \mu \geq \frac{1}{2} \) and \( \mu < \frac{1}{2} \).  
We first consider the case \( \mu \geq \frac{1}{2} \), which means \( 1 - \mu \leq \mu \). The expression of $g(z)$ is given by
\begin{equation}
g(z) =
\left\{
\begin{array}{ll}
-2\mu, & \mathrm{if }~ z \geq 1 + \mu, \\
\frac{1}{2}(z - \mu - 1)^2 - 2\mu, & \mathrm{if } ~ \mu \leq z < 1 + \mu, \\
\frac{1}{2} - z - \mu, & \mathrm{if } ~ 1 - \mu \leq z < \mu, \\
-\frac{1}{2}(z + \mu)^2, & \mathrm{if }~ -\mu \leq z < 1 - \mu, \\
0, & \mathrm{if }~ z < -\mu.
\end{array}
\right.
\end{equation}
It can be seen that when $\mu\geq 1/2$, $g(z)$ is continuous and segment-wise non-increasing.
Hence, $g(z)$ is monotonically non-increasing in this case.
We then consider the case $\mu< 1/2$ which means $1-\mu> \mu$.
The expression of $g(z)$ is given by
\begin{equation}
g(z) =
\left\{
\begin{array}{ll}
-2\mu, & \mathrm{if~} z \geq 1 + \mu, \\
\frac{1}{2}(z - \mu - 1 )^2 - 2\mu, & \mathrm{if~} 1 - \mu \leq z < 1 + \mu, \\
-2\mu z, & \mathrm{if~} \mu \leq z < 1 - \mu, \\
-\frac{1}{2}(\mu + z)^2, & \mathrm{if~} -\mu \leq z < \mu, \\
0, & \mathrm{if~} z < -\mu.
\end{array}
\right.
\end{equation}
It can be seen that $g(z)$ is also monotonically non-increasing.
\hfill $\blacksquare$
\medskip\\
Lemma \ref{lem2} implies that if there exist indices $p\in\{1,...,k\}$ and $q\in\{k+1,...,n\}$ such that $z_{j_p}<z_{j_q}$, then we can construct ${\hat{\bx}}$ as
\begin{equation}
\hat{x}_i =
\left\{
\begin{array}{ll}
x_q, & \mathrm{if~} i = p, \\
x_p, & \mathrm{if~} i = q, \\
x_i, & \mathrm{if~} i \neq p\ \mathrm{and}\ i \neq q.
\end{array}
\right.
\end{equation}
This leads to a non-increasing objective value:
\begin{equation}
\begin{array}{rl}
&\tilde{u}(\bx) - \tilde{u}(\hat{\bx})\\
= & 
\displaystyle
\left( \frac{1}{2} \left( \left[ z_{j_p} + \mu \right]_0^1 - z_{j_p} \right)^2 - \mu \left[ z_{j_p} + \mu \right]_0^1 \right) \\
& + \left( \frac{1}{2} \left( \left[ z_{j_q} - \mu \right]_0^1 - z_{j_q} \right)^2 + \mu \left[ z_{j_q} - \mu \right]_0^1 \right) \\
& - \left( \frac{1}{2} \left( \left[ z_{j_q} + \mu \right]_0^1 - z_{j_q} \right)^2 - \mu \left[ z_{j_q} + \mu \right]_0^1 \right) \\
& - \left( \frac{1}{2} \left( \left[ z_{j_p} - \mu \right]_0^1 - z_{j_p} \right)^2 + \mu \left[ z_{j_p} - \mu \right]_0^1 \right) \\
= & g(z_{j_p}) - g(z_{j_q}) \geq 0.
\end{array}
\end{equation}
Hence, based on Lemma \ref{lem2}, $\tilde u(\bx)$ can be further lower bounded as
\begin{align}
\tilde{u}(\bx) \geq 
&\sum_{i=1}^{k} \left( \frac{1}{2} \left( \left[ z_i + \mu \right]_0^1 - z_i \right)^2 - \mu \left[ z_i + \mu \right]_0^1 \right) \notag \\
&+ \sum_{i=k+1}^{n} \left( \frac{1}{2} \left( \left[ z_i - \mu \right]_0^1 - z_i \right)^2 + \mu \left[ z_i - \mu \right]_0^1 \right),
\label{eq:long_ineq}
\end{align}
which does not depend on $\bx$.
It can be verified that when 
\begin{equation}
x_i =
\left\{
\begin{array}{ll}
\left[ z_i + \mu \right]_0^1, & i \leq k, \\
\left[ z_i - \mu \right]_0^1, & i > k,
\end{array}
\right.
\label{eq:sol}
\end{equation}
the lower bound is obtained.
Hence, we can conclude that (\ref{eq:sol}) is an optimal solution to (\ref{eq:p}).

\section{C: Proof of Proposition \ref{prop_dks_prox_convergence}}
Recall that the PGM algorithm iterates as
\begin{equation}\label{eq:extro_append}
\bz^{\ell} = \bx^\ell + \gamma_\ell (\boldsymbol{x}^\ell-\boldsymbol{x}^{\ell-1}), 
\end{equation} 
\begin{equation}\label{eq:prox_append}
\bx^{\ell+1} = \text{prox}_{[0, 1]^n, \eta_\ell \lambda h}( \bz^\ell - \eta_\ell \nabla f(\bz^\ell) ),
\end{equation} 
where $\{\gamma_\ell\}_{\ell\geq 1}$ is an extrapolation sequence.
We assume that the step sizes satisfy $c_1L_f\leq 1/\eta_\ell\leq c_2L_f$ where $1<c_1\leq c_2<\infty$, and $0\leq\gamma_\ell<\bar \gamma<1$ where $\bar\gamma=(c_1-1)/(2+2c_2)$.

To begin with the proof, we define
\[
g(\bx | \tilde{\bx}, \beta) = f(\tilde{\bx}) + \langle \nabla f(\tilde{\bx}), \bx - \tilde{\bx} \rangle + \frac{\beta}{2} \| \bx - \tilde{\bx} \|_2^2.
\]
Since the objective function has $L_f$-Lipschitz continuous gradient, we have:
\begin{equation}
\label{eq:f_lip}
f(\bx^{\ell+1}) \leq f(\bx^{\ell}) + \nabla f(\bx^{\ell})^\top( \bx^{\ell+1} - \bx^{\ell}) + \frac{L_f}{2} \| \bx^{\ell+1} - \bx^{\ell} \|_2^2.    
\end{equation}
Then, we can rewrite \eqref{eq:prox_append} as
\begin{equation}
\begin{aligned}
\bx^{\ell+1} \in \arg \min_{\bx \in [0,1]^n} g(\bx | \bz^\ell, 1 / \eta_\ell) + \lambda  h(\bx).
\label{eq_60}
\end{aligned}
\end{equation}
Since \(\bx^{\ell+1}\) is a critical point of the problem in \eqref{eq_60}, \(\bx^{\ell+1}\) satisfies
\begin{equation}\label{eq:xlp1_opt}
\begin{aligned}
\boldsymbol{0} &  \in \nabla g(\bx^{\ell+1}| \bz^\ell, 1/\eta_\ell) + \lambda \partial h(\bx^{\ell+1}),\\
&  = \nabla f(\bz^{\ell}) + (\bx^{\ell+1} - \bz^\ell)/\eta_\ell + \lambda \partial   h(\bx^{\ell+1}).
\end{aligned}
\end{equation}
It follows that
\begin{equation}
    \begin{aligned}
    &\text{dist}(\boldsymbol{0}, \partial F_\lambda(\bx^{\ell+1}))^2  \\=& \text{dist}(\boldsymbol{0}, \nabla f(\bx^{\ell+1}) + \lambda \partial   h(\bx^{\ell+1}))^2 \\
	=& \inf_{\bv \in\partial h(\bx^{\ell+1})} \Arrowvert \nabla f(\bx^{\ell+1}) + \bv\Arrowvert_2^2\\
	\leq& \left\|\nabla f(\bx^{\ell+1}) - \left(\nabla f(\bz^\ell) + (\bx^{\ell+1} - \bz^\ell)/\eta_\ell \right)\right\|_2^2   \\
	 \leq & 2(L_f^2+\eta_\ell^{-2}) \| \bx^{\ell+1} - \bz^\ell \|_2^2,
    \end{aligned}
\end{equation}
where we recall that the distance is defined as
\[
\mathrm{dist}(\bx, \mathcal{A}) := \inf_{\by \in \mathcal{A}} \| \bx - \by \|_2.
\]
Hence, we have the result
\begin{equation}\label{eq:dist_upperbound}
    \begin{aligned}
     &\text{dist}(\boldsymbol{0}, \partial F_\lambda(\bx^{\ell+1})) \\
    \leq & \sqrt{2(L_f^2+\eta_\ell^{-2})} \| \bx^{\ell+1} - \bx^\ell-\gamma_\ell(\bx^\ell-\bx^{\ell-1}) \|_2,
    \\ \leq &C_1\left(\| \bx^{\ell+1} - \bx^\ell \|_2+\| \bx^{\ell} - \bx^{\ell-1} \|_2\right),
    \end{aligned}
\end{equation}
where $C_1=\sqrt{2(1+c_2^2)}L_f$.

On the other hand,
we see from \eqref{eq_60} that
\begin{equation}\label{eq:xlp1_xl}
\begin{aligned}
&g(\bx^{\ell+1} \mid \bz^\ell, 1/\eta_\ell) + \lambda h(\bx^{\ell+1}) 
\\\leq& g(\bx^\ell \mid \bz^\ell, 1 / \eta_\ell) + \lambda h(\bx^\ell).
\end{aligned}
\end{equation}
Adding \eqref{eq:f_lip} and \eqref{eq:xlp1_xl} we get
\begin{equation}\label{eq:F-F}
\begin{aligned}
    &F_{\lambda}(\bx^\ell)-F_{\lambda}(\bx^{\ell+1})\\
    \geq&\left(\nabla f(\bz^\ell)-\nabla f(\bx^\ell)\right)^\top(\bx^{\ell+1}-\bx^{\ell})+\frac{1}{2\eta_\ell}\|\bx^{\ell+1}-\bz^\ell\|_2^2\\
    &-\frac{L_f}{2}\|\bx^{\ell+1}-\bx^{\ell}\|_2^2-\frac{1}{2\eta_\ell}\|\bx^{\ell}-\bz^\ell\|_2^2\\
    =&\left(\nabla f(\bz^\ell)-\nabla f(\bx^\ell)\right)^\top(\bx^{\ell+1}-\bx^{\ell})\\
    &+\frac{1-\eta_\ell L_f}{2\eta_\ell}\|\bx^{\ell+1}-\bx^{\ell}\|_2^2+\frac{1}{\eta_\ell}(\bx^{\ell+1}-\bx^{\ell})^\top(\bx^{\ell}-\bz^{\ell})\\
    \geq & -\left(L_f+\frac{1}{\eta_\ell}\right)\|\bx^{\ell+1}-\bx^{\ell}\|_2\|\bx^{\ell}-\bz^{\ell}\|_2\\
    &+\frac{1-\eta_\ell L_f}{2\eta_\ell}\|\bx^{\ell+1}-\bx^{\ell}\|_2^2\\
    =&-\gamma_l\left(L_f+\frac{1}{\eta_\ell}\right)\|\bx^{\ell+1}-\bx^{\ell}\|_2\|\bx^{\ell}-\bx^{\ell-1}\|_2\\
    &+\frac{1-\eta_\ell L_f}{2\eta_\ell}\|\bx^{\ell+1}-\bx^{\ell}\|_2^2\\
    \geq &d_\ell\|\bx^{\ell+1}-\bx^{\ell}\|_2^2 - e_\ell\|\bx^{\ell}-\bx^{\ell-1}\|_2^2,
\end{aligned}
\end{equation}
where
\[
d_\ell =\frac{1/\eta_\ell-L_f}{4}>0,\ e_\ell = \frac{\gamma_\ell^2(L_f+1/\eta_\ell)^2}{1/\eta_\ell-L_f}>0;
\]
we used the Cauchy–Schwarz inequality and the Lipschitz continuity of $\nabla f(\bx)$ in the second inequality; in the last inequality, we used Young's inequality
\(ab\leq \frac{a^2}{2\varepsilon}+\frac{b^2\varepsilon}{2}, \forall a, b\in\mathbb{R},\ \forall \varepsilon\in\mathbb{R}_+
\)
with
\(a = (L_f+1/\eta_\ell)\gamma_\ell\|\bx^\ell-\bx^{\ell-1}\|_2\), \(b=\|\bx^{\ell+1}-\bx^{\ell}\|_2\), and \(\varepsilon =(1/\eta_\ell-L_f)/2\).
Note that the formula \eqref{eq:F-F} is an adaptation of Lemma 1 in \cite{xu2017globally}.
As a result, we have
\begin{equation}
    \begin{aligned}
    &F_{\lambda}(\bx^0)-F_{\lambda}^\star\\
        \geq &F_{\lambda}(\bx^0)-F_{\lambda}(\bx^{J+1})\\
        =&\sum_{\ell=0}^{J}F_{\lambda}(\bx^\ell)-F_{\lambda}(\bx^{\ell+1}) \\
        \geq & \sum_{\ell=0}^{J}d_l\|\bx^{\ell+1}-\bx^{\ell}\|_2^2 - e_l\|\bx^{{\ell}}-\bx^{\ell-1}\|_2^2\\
        =&d_\ell\|\bx^{\ell+1}-\bx^{\ell}\|_2^2+\sum_{\ell=0}^{J-1}(d_l-e_{\ell+1})\|\bx^{\ell+1}-\bx^{\ell}\|_2^2\\
        \geq &\sum_{\ell=0}^{J}(d_l-e_{\ell+1})\|\bx^{\ell+1}-\bx^{\ell}\|_2^2\\
        \geq & \frac{C_2(J+1)}{2}\min_{\ell=0,...,J} \|\bx^{\ell+1}-\bx^{\ell}\|_2^2 + \|\bx^{\ell}-\bx^{\ell-1}\|_2^2.
    \end{aligned}
\end{equation}
where
\[
\begin{aligned}
C_2 &=\min_{l=0,...,\ell} d_l-e_{l+1}\\
&= \left(\frac{c_1-1}{4}-\frac{\bar\gamma^2( c_2+1)^2}{c_1-1}\right)L_f>0. 
\end{aligned}
\]
By using $a+b\leq \sqrt{2a^2+b^2}$, we get
\begin{equation}\label{eq:sqrt_dist}
    \begin{aligned}
        &\min_{l=0,...,J} \|\bx^{l+1}-\bx^{l}\|_2 + \|\bx^{l}-\bx^{l-1}\|_2\\
        \leq & \sqrt{\frac{4(F_{\lambda}(\bx^0)-F_{\lambda}^\star)}{C_2(J+1)}}.
    \end{aligned}
\end{equation}
Combining \eqref{eq:sqrt_dist} and \eqref{eq:dist_upperbound}, we finally obtain the sublinear convergence rate:
\begin{equation}
    \min_{l=0,...,J} \text{dist}(\boldsymbol{0}, \partial F_\lambda(\bx^{\ell+1}))\leq C_1 \sqrt{\frac{4(F_{\lambda}(\bx^0)-F_{\lambda}^\star)}{C_2(J+1)}}.
\end{equation}

\section{D: Lipschitz Constants}
Consider the function $f(\bx) = -\bx^\top\bA\bx$. 
The following holds:
\begin{itemize}
    \item [i)] $\nabla f(\bx)$ is $2\| \bA\|_2$-Lipschitz continuous on $\mathbb{R}^n$;
    \item [ii)] $f(\bx)$ is $2 \sqrt{k}\|\bA\|_2$-Lipschitz continuous on $\text{conv}(\setU^n_k)$;
    \item[iii)] $f(\bx)$ is $2 \sqrt{n}\| \bA\|_2$-Lipschitz continuous on $[0,1]^n$.
\end{itemize}

{\em Proof of}\  i): 
Note that $\nabla f(\bx) = -2\bA\bx$.
For any $\bx$, $\by\in \mathbb{R}^{n}$, we have
\begin{equation}
\begin{aligned}
     \|  \nabla {f}(\bx) - \nabla{f}(\by) \|_2 
     & =\| -2\bA\bx + 2\bA \by \|_2
     \\& = \| 2\bA  (\bx-\by)\|_2 
    \\& \leq 2 \Arrowvert \bA\Arrowvert_2 \Arrowvert\bx-\by\Arrowvert_2.
\end{aligned}
\end{equation}
Therefore, the Lipschitz constant is $2 \Arrowvert \bA\Arrowvert_2$.
\hfill 
$\blacksquare$
\medskip

{\em Proof of}\  ii): 
For any $\bx$, $\by\in\text{conv}(\setU^n_k)$, we have:
\begin{equation}
\begin{aligned}
     \arrowvert  {f}(\bx) - {f}(\by) \arrowvert 
     & =\arrowvert \bx^\top {\bA}  \bx -\by^\top{\bA}  \by \arrowvert
     \\& = \arrowvert (\bx+\by)^\top {\bA}  (\bx-\by)\arrowvert 
    \\& \leq \Arrowvert \bx+\by\Arrowvert_2\Arrowvert {\bA} \Arrowvert_2\Arrowvert\bx-\by\Arrowvert_2
    \\&
    \leq2\sqrt{k} \Arrowvert {\bA}  \Arrowvert_2 \Arrowvert \bx-\by\Arrowvert_2,
\end{aligned}
\end{equation}
where we used the fact that $ \Arrowvert\bx+\by\Arrowvert_2\leq 2\sqrt{k}$. 
The Lipschitz constant is $2\sqrt{k} \Arrowvert {\bA}  \Arrowvert_2$.
\hfill 
$\blacksquare$

{\em Proof of}\ iii): 
For any $\bx$, $\by\in[0, 1]^n$, we have:
\begin{equation}
\begin{aligned}
     \arrowvert  {f}(\bx) - {f}(\by) \arrowvert 
    & \leq \Arrowvert \bx+\by\Arrowvert_2\Arrowvert {\bA} \Arrowvert_2\Arrowvert\bx-\by\Arrowvert_2
    \\&
    \leq2\sqrt{n} \Arrowvert {\bA}  \Arrowvert_2 \Arrowvert \bx-\by\Arrowvert_2,
\end{aligned}
\end{equation}
where we used the fact that $ \Arrowvert\bx+\by\Arrowvert_2\leq 2\sqrt{n} $. 
The Lipschitz constant is $2\sqrt{n} \Arrowvert {\bA}  \Arrowvert_2$.
\hfill $\blacksquare$

\section{E: Formulation of the D$k_1k_2$BS Problem \eqref{eq_dkbs}}
\label{appendix_dkbs}
Consider the D$k_1k_2$BS Problem \eqref{eq_dkbs}.
Let $\ba = [\bx^\top \; \by^\top]^\top \in \mathbb{R}^n$ where $n = n_1 + n_2$, and define the symmetric adjacency matrix:
\begin{equation}
\bA =
\left[
\begin{array}{cc}
\boldsymbol{0} & \bB \\
\bB^\top & \boldsymbol{0}
\end{array}
\right]
\in \mathbb{R}^{n \times n}.
\end{equation}
It follows that $\bx^\top \bB \by = \frac{1}{2} \ba^\top \bA \ba$.
The feasible set can be rewritten as:
\begin{equation}
\setU_{k_1, k_2}^{n_1, n_2} := 
\left\{ 
\ba = 
\left[
\begin{array}{c}
\bx \\
\by
\end{array}
\right]
\in \{0,1\}^n \;\big|\; \bx \in \setU_{k_1}^{n_1},\; \by \in \setU_{k_2}^{n_2} 
\right\}.
\end{equation}
Then, the D$k_1k_2$BS problem~(\ref{eq_dkbs}) is equivalent to the following quadratic minimization problem:
\begin{equation}
\min_{\ba \in \setU_{k_1, k_2}^{n_1, n_2}} \quad f(\ba) := - \ba^\top \bA \ba.
\label{eq_dkbs_z}
\end{equation}
The difference to the D$k$S problem is that the above problem has two cardinality constraints on two disjoint segments of the variable $\ba$.
There is no need to treat the two vector variables $\bx$ and $\by$ separately;
our error bound function and the proposed PGM algorithm can be directly applied with the variable $\ba$.

\section{F: More Descriptions and  Results for the D$k_1k_2$BS Problem}
We preprocess all the datasets by converting each graph into a simple, undirected form by removing edge directions (if present), eliminating multi-edges, and discarding isolated nodes. Furthermore, all nonzero edge weights are set to $1$.
We follow the implementation in Algorithm~\ref{alg_whole}.
All the methods are initialized with the vector $\ba^0 =\bone/(k_1+k_2)$.
The initial penalty parameter is set to \( \lambda_0 = 10^{-10} \). The whole algorithm terminates when either \( \| \ba^{\ell+1} - \ba^\ell \|_2^2 \leq 10^{-15} \), or the number of iterations exceeds $100$. The penalty parameter \( \lambda \) is updated by \( \lambda_{\ell+1} = 10 \lambda_\ell \) when either \( \| \ba^{\ell+1} - \ba^\ell \|_2 / \| \ba^{\ell+1} \|_2 < 0.5 \) or $10$ iterations have passed since the last update.

For the experimental evaluation, we also consider varying the value of $k_1$ from $10$ to $1000$, and set $k_2=k_1$.
The corresponding results of edge density and runtime are shown in Figure \ref{fig_dkbs_density} and \ref{fig_dkbs_time}.
The proposed method, EP-Prox, consistently outperforms Rank-1 PC and demonstrates competitive performance compared to EXPP, while being faster than EXPP and a little slower than Rank-1 PC.

\begin{figure}[!t]
\centering
\begin{minipage}[b]{0.48\columnwidth}
\centering
\includegraphics[width=\linewidth]{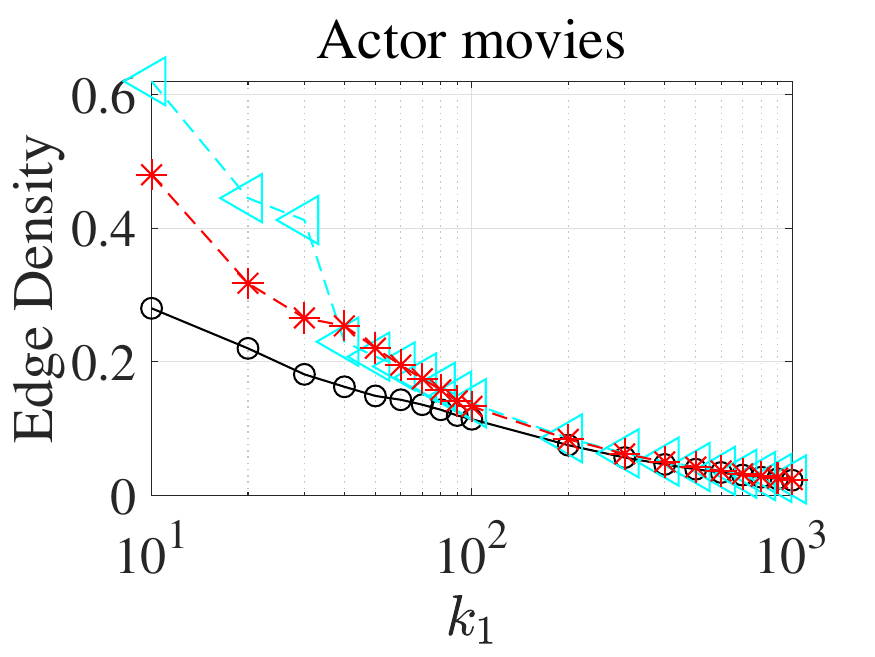}
\end{minipage}
\hfill
\begin{minipage}[b]{0.48\columnwidth}
\centering
\includegraphics[width=\linewidth]{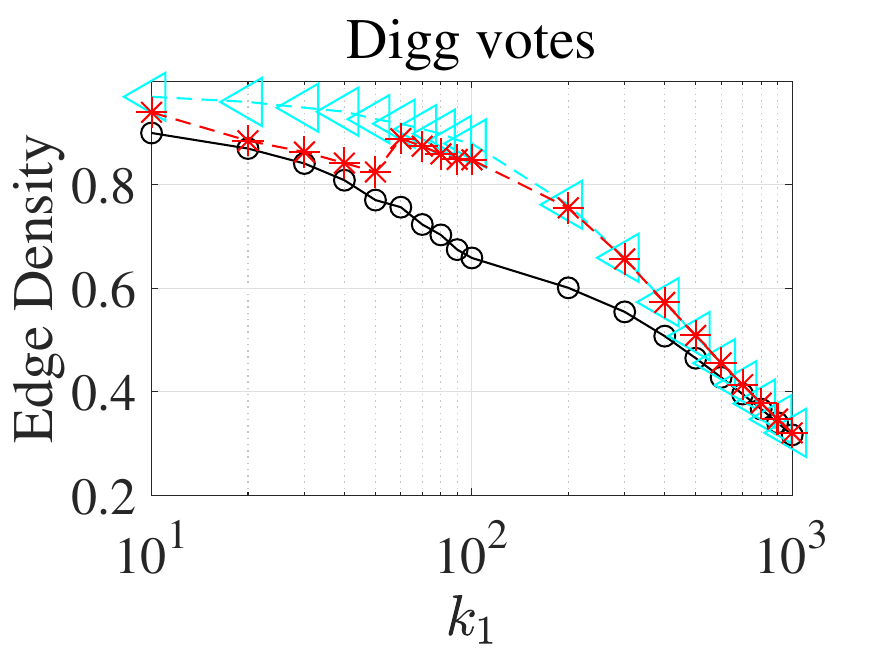}
\end{minipage}
\hfill
\begin{minipage}[b]{0.48\columnwidth}
\centering
\includegraphics[width=\linewidth]{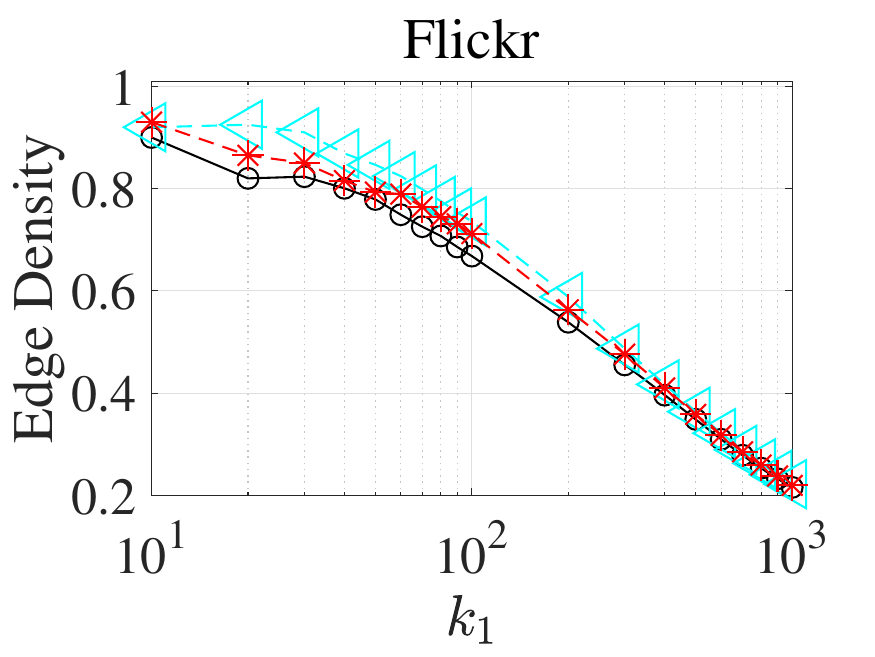}
\end{minipage}
\hfill
\begin{minipage}[b]{0.48\columnwidth}
\centering
\includegraphics[width=\linewidth]{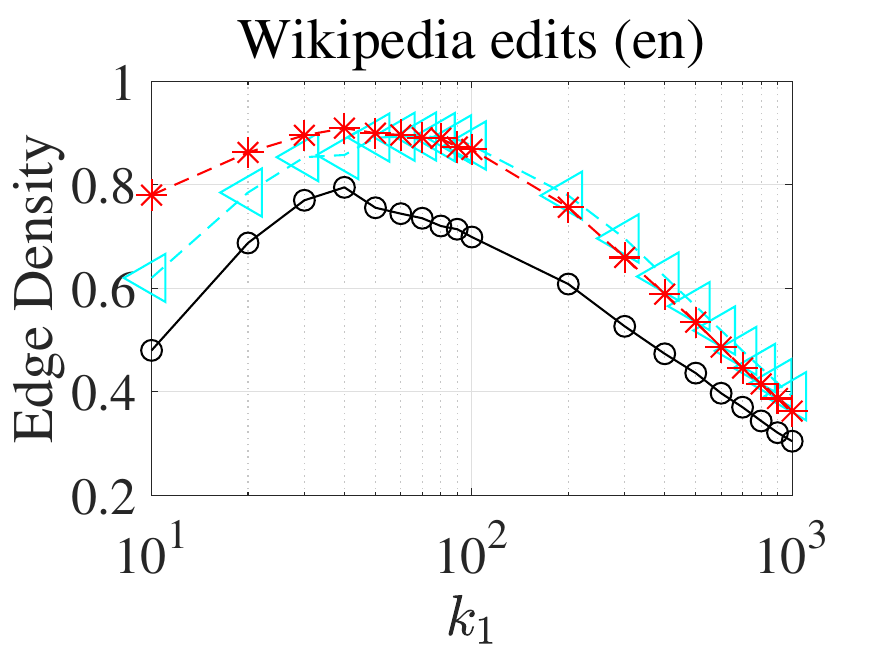}
\end{minipage}
\hfill
\begin{minipage}[b]{0.99\columnwidth}
\centering
\includegraphics[width=0.45\linewidth]{figures/dkbs/dkbs_legend.jpg}
\end{minipage}
\caption{Edge density under different \( k_1 = k_2 \) for D$k_1k_2$BS.}
\label{fig_dkbs_density}
\end{figure}

\begin{figure}[H]
\centering
\begin{minipage}[b]{0.48\columnwidth}
\centering
\includegraphics[width=\linewidth]{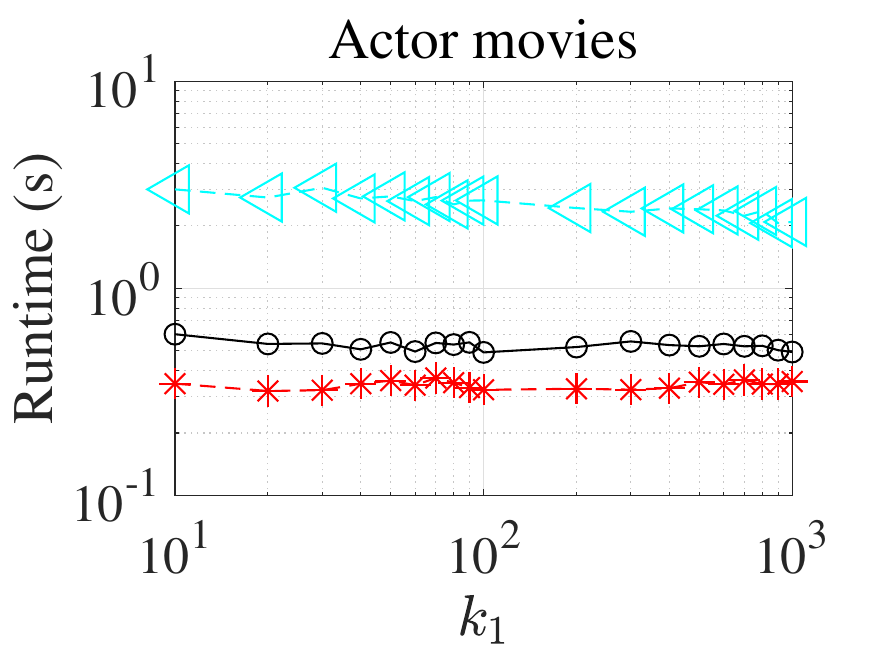}
\end{minipage}
\hfill
\begin{minipage}[b]{0.48\columnwidth}
\centering
\includegraphics[width=\linewidth]{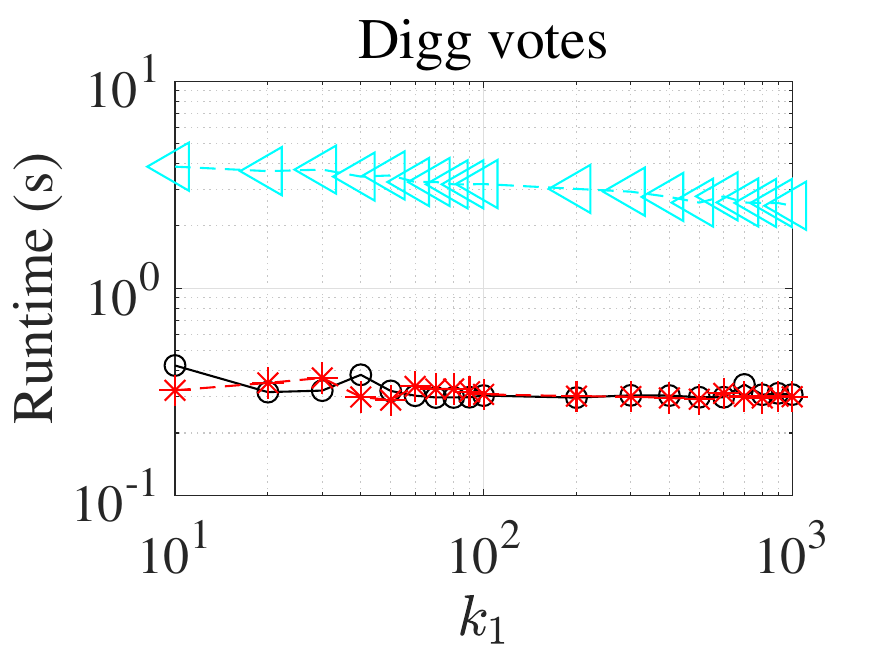}
\end{minipage}
\hfill
\begin{minipage}[b]{0.48\columnwidth}
\centering
\includegraphics[width=\linewidth]{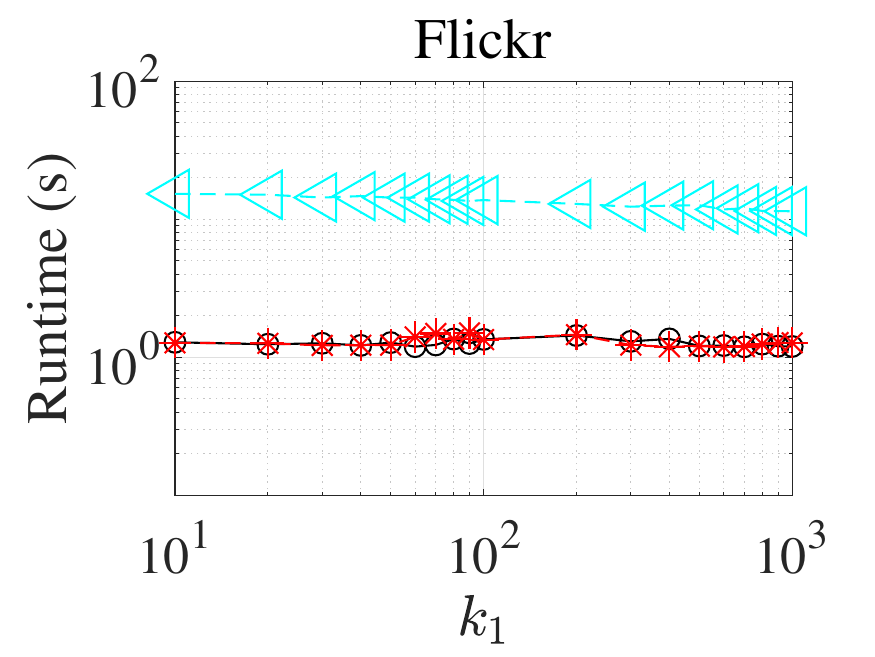}
\end{minipage}
\hfill
\begin{minipage}[b]{0.48\columnwidth}
\centering
\includegraphics[width=\linewidth]{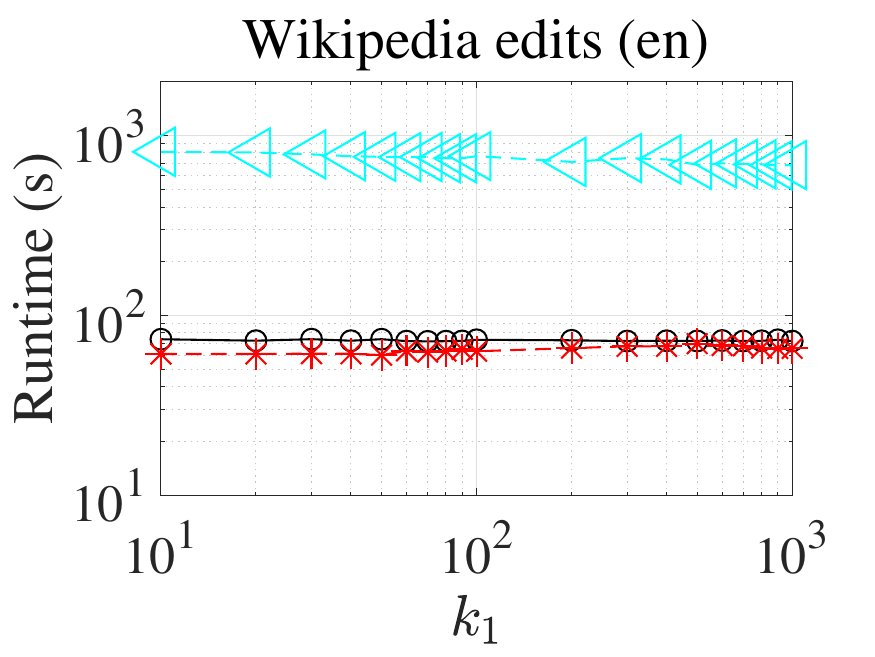}
\end{minipage}
\hfill
\begin{minipage}[b]{0.99\columnwidth}
\centering
\includegraphics[width=0.45\linewidth]{figures/dkbs/dkbs_legend.jpg}
\end{minipage}
\caption{Runtime under different \( k_1 = k_2 \) for D$k_1k_2$BS.}
\label{fig_dkbs_time}
\end{figure}

\end{document}